\documentclass{emulateapj}
\usepackage{graphicx}
\usepackage{amssymb}
\usepackage{amsmath}
\usepackage{natbib}

\newcommand{\ha}{H$\alpha$}
\newcommand{\hb}{H$\beta$}

\def\ltsima{$\buildrel<\over\sim$}
\def\lsim{\lower.5ex\hbox{\ltsima}~}
\def\gtsima{$\buildrel>\over\sim$}
\def\gsim{\lower.5ex\hbox{\gtsima}~}

\def\msolyr{M$_{\odot}$~yr$^{-1}$}
\def\msol{M$_{\odot}$}
\def\mstar{M$_{\star}$}

\def\teff{\ifmmode T_{\rm eff} \else $T_{\mathrm{eff}}$\fi}

\def\ha{H$\alpha$} 
\def\hb{H$\beta$}

\def\cm2{cm$^{-2}$}

\def\ewha{$EW_{\mathrm{H}\alpha}$}
\def\hi{H{\sc i}}

\def\oiii{O{\sc iii}}
\def\oii{O{\sc ii}}
\def\nii{N{\sc ii}}
\def\sii{S{\sc ii}}
\def\nh{\ifmmode N_{\mathrm{HI}}\else $N_{\mathrm{HI}}$\fi}

\def\vexp{\ifmmode v_{\rm exp} \else v$_{\rm exp}$\fi}
\def\taua{\ifmmode \tau_{a}\else $\tau_{a}$\fi}

\begin{document} 

\title{Hubble Space Telescope Grism Spectroscopy of Extreme Starbursts Across Cosmic Time: The Role of Dwarf Galaxies in the Star Formation History of the Universe\footnotemark[$\dagger$]}

\footnotetext[$\dagger$]{Based on observations made with the NASA/ESA Hubble Space Telescope, which is operated by the Association of Universities for Research in Astronomy, Inc., under NASA contract NAS 5-26555. These observations are associated with programs 11696, 12283, 12568, 12177, and 12328.}

\author{Hakim Atek\altaffilmark{1,2}, 
Jean-Paul Kneib\altaffilmark{1,3},
Camilla Pacifici\altaffilmark{4},
Matthew Malkan\altaffilmark{5},
Stephane Charlot\altaffilmark{6},
Janice Lee\altaffilmark{7},
Alejandro Bedregal \altaffilmark{8},
Andrew J. Bunker \altaffilmark{9}, 
James W. Colbert\altaffilmark{2},
Alan Dressler\altaffilmark{10},
Nimish Hathi\altaffilmark{3},
Matthew Lehnert\altaffilmark{6},
Crystal L. Martin\altaffilmark{11},
Patrick McCarthy\altaffilmark{10},
Marc Rafelski\altaffilmark{2},
Nathaniel Ross\altaffilmark{5},
Brian Siana\altaffilmark{12},
Harry I. Teplitz\altaffilmark{13} 
}

\altaffiltext{1}{Laboratoire d'Astrophysique, EPFL, CH-1290 Sauverny, Switzerland}
\altaffiltext{2}{Spitzer Science Center, Caltech, Pasadena, CA 91125, USA}
\altaffiltext{3}{Aix Marseille Universit\'e, CNRS, LAM (Laboratoire d'Astrophysique de Marseille) UMR 7326, 13388, Marseille, France}
\altaffiltext{4}{Yonsei University Observatory, Yonsei University, Seoul 120-749, Republic of Korea}
\altaffiltext{5}{Department of Physics and Astronomy, University of California, Los Angeles, CA, USA}
\altaffiltext{6}{UPMC-CNRS, UMR7095, Institut d'Astrophysique de Paris, F-75014 Paris, France}
\altaffiltext{7}{Space Telescope Science Institute, 3700 San Martin Drive, Baltimore, MD 21218, USA}
\altaffiltext{8}{Minnesota Institute for Astrophysics, University of Minnesota, Minneapolis, MN 55455, USA}
\altaffiltext{9}{Department of Physics, University of Oxford, Denys Wilkinson Building, Keble Road, OX13RH, U.K.}
\altaffiltext{10}{Observatories of the Carnegie Institution for Science, Pasadena, CA 91101, USA}
\altaffiltext{11}{Dep't. of Physics, Univ. of Calif. Santa Barbara, CA 93106, USA}
\altaffiltext{12}{Department of Physics and Astronomy, University of California Riverside, Riverside, CA 92521, USA}
\altaffiltext{13}{Infrared Processing and Analysis Center, Caltech, Pasadena, CA 91125, USA}

\begin{abstract}
Near infrared slitless spectroscopy with the Wide Field Camera 3, onboard the {\em Hubble Space Telescope}, offers a unique opportunity to study low-mass galaxy populations at high-redshift ($z \sim$ 1-2). While most high$-z$ surveys are biased towards massive galaxies, we are able to select sources via their emission lines that have very-faint continua. We investigate the star formation rate (SFR)-stellar mass (\mstar) relation for about 1000 emission-line galaxies identified over a wide redshift range of $0.3 \lesssim z \lesssim 2.3$. We use the \ha\ emission as an accurate SFR indicator and correct the broadband photometry for the strong nebular contribution to derive accurate stellar masses down to \mstar\ $\sim 10^{7}$ \msol. We focus here on a subsample of galaxies that show extremely strong emission lines (EELGs) with rest-frame equivalent widths ranging from 200 to 1500 \AA. This population consists of outliers to the normal SFR-\mstar\ sequence with much higher specific SFRs ($> 10$ Gyr$^{-1}$). While on-sequence galaxies follow continuous star formation processes, EELGs are thought to be caught during an extreme burst of star formation that can double their stellar mass in a period of less than $100$ Myr. The contribution of the starburst population to the total star formation density appears to be larger than what has been reported for more massive galaxies in previous studies. In the complete mass range $8.2 <$ log(\mstar/\msol) $< 10$ and a SFR lower completeness limit of about 2 \msolyr (10 \msolyr) at $z \sim 1$ ($z \sim 2$), we find that starbursts having EW$_{rest}$(\ha) $>$ 300, 200, and 100 \AA\ contribute up to $\sim 13$, 18, and 34\%, respectively, to the total SFR of emission-line selected sample at $z \sim 1-2$. The comparison with samples of massive galaxies shows an increase in the contribution of starbursts towards lower masses.

\end{abstract}

\maketitle

\section{Introduction}
\label{sec:intro}

The formation and evolution of galaxies is governed by the complex interplay of key processes that include galaxy mergers, cold gas accretion feeding star formation, and the metal-enriched gas outflows driven by supernovae and supermassive black holes \citep{dekel09,bouche10,silk13}. It is now well established that the star formation history of the Universe reached its peak around $z \sim 2$ and formed most of its stellar mass by $z \sim 1$ \citep{madau96,hopkins04}. During the last decade, a correlation between the star formation rate (SFR) and the stellar mass (\mstar) has been explored up to those redshifts \citep{brinchmann04,peng10,karim11,hathi13}. Normal star-forming galaxies, lying on the so-called SFR main sequence, may be a natural consequence of smooth gas accretion, while outliers above this sequence undergo starburst episodes, which are likely driven by galaxy interactions. However, the slope and the dispersion of the SFR-\mstar\ relation are hardly consistent from one study to another \citep{daddi07,wuyts11,whitaker12}.

Most of the uncertainties may reside in the selection techniques, the difficulty to obtain a well calibrated SFR indicator at high redshift, and the use of different indicators at different redshifts. 

First, while local studies use the \ha\ emission line as an accurate measurement of the current SFR, high-redshift studies (at $z>0.5$) must rely on continuum-based calibrations, such as spectral energy distribution (SED) models or ultraviolet (UV) + far infrared (FIR), since \ha\ shifts to NIR wavelengths. 

Secondly, most of the samples assembled at high redshift are biased towards bright, hence massive, galaxies of log(\mstar/\msol) $\gtrsim10$ \citep[e.g.,][]{bauer11,whitaker12} and will likely miss the star formation occurring in faint-continuum systems \citep{Alavi13}. Therefore, to obtain a comprehensive picture of galaxy formation and evolution, one needs to account for the low-mass galaxies, which might experience a different mode of star-formation compared to the massive galaxies of the main sequence. The differences in the two populations that might influence their star formation histories (SFH) include the dynamical time scale, the feedback efficiency, and the morphology, which appears to regulate star formation more efficiently in massive galaxies \citep[e.g.,][]{lee07}.

In this context, Wide Field Camera 3 (WFC3) NIR grisms onboard {\em HST} offer a unique opportunity to select faint galaxies by their emission lines over a wide redshift range $ 0.3 \lesssim z \lesssim 2.3$. In \citet{atek11}, we unveiled a population of extremely strong emission-line galaxies with equivalent widths up to 1000 \AA\ and stellar masses as low as \mstar $\sim 10^{7}$ \msol. While extremely high EW galaxies are relatively rare in the local universe \citep{cardamone09, amorin10}, their number density increases by an order of magnitude or more at $z \sim 2-4$ \citep{kakazu07,atek11, shim11}. Low-mass starburst galaxies offer a complementary approach to the study of the star formation history of the Universe by exploring a different mode of mass assembly over different time scales compared to their massive counterparts. 

We present here a large sample of galaxies selected from {\em HST} grism spectroscopic surveys to shed new light on the SFR-\mstar\ relation and the contribution of high-$z$ dwarf galaxies to the total star formation density. The paper is structured as follows. In Section \ref{sec:data} we describe the observations, data reduction, and emission line measurements. The sample selection is presented in Section \ref{sec:selection}. In Section \ref{sec:sed} we present the stellar population modeling. We estimate the AGN contribution to our sample in Section \ref{sec:agn}. In Section \ref{sec:sfr_mass} we present the results of the SFR-\mstar\ relation, while Section \ref{sec:eelg} is devoted to the particular case of extreme emission line galaxies. We finally summarize our results in Section \ref{sec:conclusion}. Throughout the paper, we use a standard $\Lambda$CDM cosmology with $H_0=70$\ km s$^{-1}$\ Mpc$^{-1}$, $\Omega_{\Lambda}=0.7,$\ and $\Omega_{m}=0.3$.

\vspace{1cm}

\section{Observations and Data Reduction}
\label{sec:data}

\subsection{The WISP Survey}
The WISP (WFC3 Infrared Spectroscopic Parallel) survey \citep[PI=Malkan;][]{atek10} is a large pure parallel program using the near-IR (NIR) grism capabilities of the Wide Field Camera 3 (WFC3) onboard {\em HST} to observe a large number of uncorrelated fields over five {\em HST} cycles (GO-11696, GO-12283, GO-12568, GO-12902 \& GO-13352). Typically, data consist of slitless spectroscopy in both NIR grisms G102 ($0.8-1.2$ \micron, R$\sim 210$) and G141 ($1.2-1.7$ \micron, R$\sim 130$), whereas in the case of short visits only G141 is used. This is complemented by NIR imaging in F110W and F160W, and UVIS imaging in F475X and F600LP for the longer opportunities. The IR and UVIS channels have a plate scale of 0\farcs13 and 0\farcs04 for a total field of view of 123\arcsec$\times$136\arcsec and 162\arcsec$\times$162\arcsec, respectively. In addition, a follow-up program with the {\em Spitzer Space Telescope} \citep[][PI=Colbert; GO-80134, GO-90230]{werner04} provides IRAC 3.6\micron\ imaging for a subset of the observed fields in WISP that have both G102 and G141 data.

The reduction of NIR data is performed using the WISP pipeline described in \citet{atek10}. For the emission line extraction we first used an automated procedure to identify the emission lines and assign redshifts. The algorithm is described in detail in \citet{colbert13}. Briefly, the program fits a cubic spline to the continuum and subtracts it from the spectrum. By dividing this spectrum by the error spectrum we obtain a signal-to-noise spectrum. Then, an emission line needs to satisfy a signal to noise ratio higher than $\sqrt{3}$ in at least three contiguous pixels. Every emission-line candidate is then independently examined in the 1D spectrum and the 2D dispersed image by two team members and automatically fitted upon confirmation. In the present study, we used the same emission line catalog presented in \citet{colbert13} consisting of 29 WISP fields which are covered by both G102 and G141 grism observations.

\subsection{The 3DHST Survey}
The second part of our sample is based on data from the 3DHST program \citep[GO 12177 \& 12328][]{brammer12}. The survey targets four well-studied fields, following up the imaging campaign of the Cosmic Assembly Near-infrared Deep Extragalactic Legacy Survey \citep[CANDELS, ][]{grogin11, koekemoer11}. Observations consist of 248 {\em HST} orbits distributed between G141 grism and F140W imaging over Cycle 18 and 19 in the EGS/AEGIS, COSMOS, GOODS-South and UKIDSS/UDS fields. 

The WISP survey is executed in pure parallel mode, which constrain the observations at a fixed orientation and position. The 3DHST survey is a primary program with the observations taken at a fixed orientation but with a dithering pattern to mitigate the detector artifacts, which was not possible with parallel observations. Another important difference is the use of only on grism (G141) instead of both in WISP (G102 and G141). The direct implication is a smaller redshift coverage for 3DHST starting at $z \sim 0.8$ compared to WISP which starts at $z \sim 0.35$. The larger spectral coverage of WISP allows us to detect multiple emission lines -mostly \ha\ and [\oiii] pairs, which provides more secure redshifts. The availability of multiple lines also makes possible the study of galaxy physical properties such as the dust extinction \citep{dominguez13} or metallicity \citet{henry13b}. The exposure time is split between the two IR grisms for WISP, while the total exposure time is devoted to the G141 grism in 3DHST data. The mean exposure time in the WISP survey is $\sim$ 2,000 s in G141 and $\sim$ 5,000 s in G102, with few deeper fields observed for up to 28,000 s.

The reduction of 3DHST data is performed with a modified version of the WISP pipeline, independently of \citet{brammer12}. Because of the large variations in the IR  sky background, we have produced a master background image for each of the four main fields using all the G141 pointings in that field. We followed the same procedure as for WISP where we masked the spectral traces in all the frames before median combining them. The median image was then cleaned by interpolation using a bad pixel mask. Finally, the master sky template is scaled to each individual grism frame during the reduction process. This new version of the pipeline also deals with dithered frames to correct for bad pixels and cosmic rays, making the cleaning procedure more efficient. Detailed description of the reduction and spectral extraction pipeline can be found in \citet{atek10}. 

Dedicated and interactive software is developed for line emission detection and flux measurements. The routine displays the 1D and 2D spectra where the user has the ability to identify one or multiple emission lines. Based on this redshift information, the program fits the continuum with a polynomial and all the emission lines within the wavelength coverage of the spectrum using one or multiple gaussians in the case of the [\oiii] doublet for instance. The central wavelength of the line is allowed to vary according to the redshift uncertainty \citep[see][]{colbert13} and the ratio [\oiii]$\lambda$5007/[\oiii]$\lambda$4959 is fixed to 3.2 \citep{osterbrock89}. The user then has the possibility to confirm each line, store the line measurements after inspecting the 1D and 2D spectra, the contamination and the significance of the line, and also assign a quality flag. Four values are possible: 1) very good (multiple emission line of high significance), 2: good (multiple lines with less significance), 3: uncertain (single line), 4: strong contamination from nearby objects. The errors on the emission line parameters are estimated using Monte Carlo simulations. We create 100 spectra by randomly adding noise within the uncertainty range of the flux density and apply the same automatic fitting procedure. The program stores the 1-$\sigma$ uncertainties derived from the probability distribution for all the parameters. In addition to the signal-to-noise ratio, the uncertainties on the line flux depend on the equivalent width because the line detection will be more difficult in the case of a bright continuum  For both survey datasets, the equivalent width completeness limit is about 30 \AA\ \citep[cf.][]{colbert13}. The average exposure time of the 3DHST is about 4500 s in the G141 grism, compared to the WISP data that use both G102 and G141 grisms with typical exposure times of 6000 s and 2000 s, respectively.

\begin{figure*}[!htbp]
   \centering
       \includegraphics[width=8.9cm]{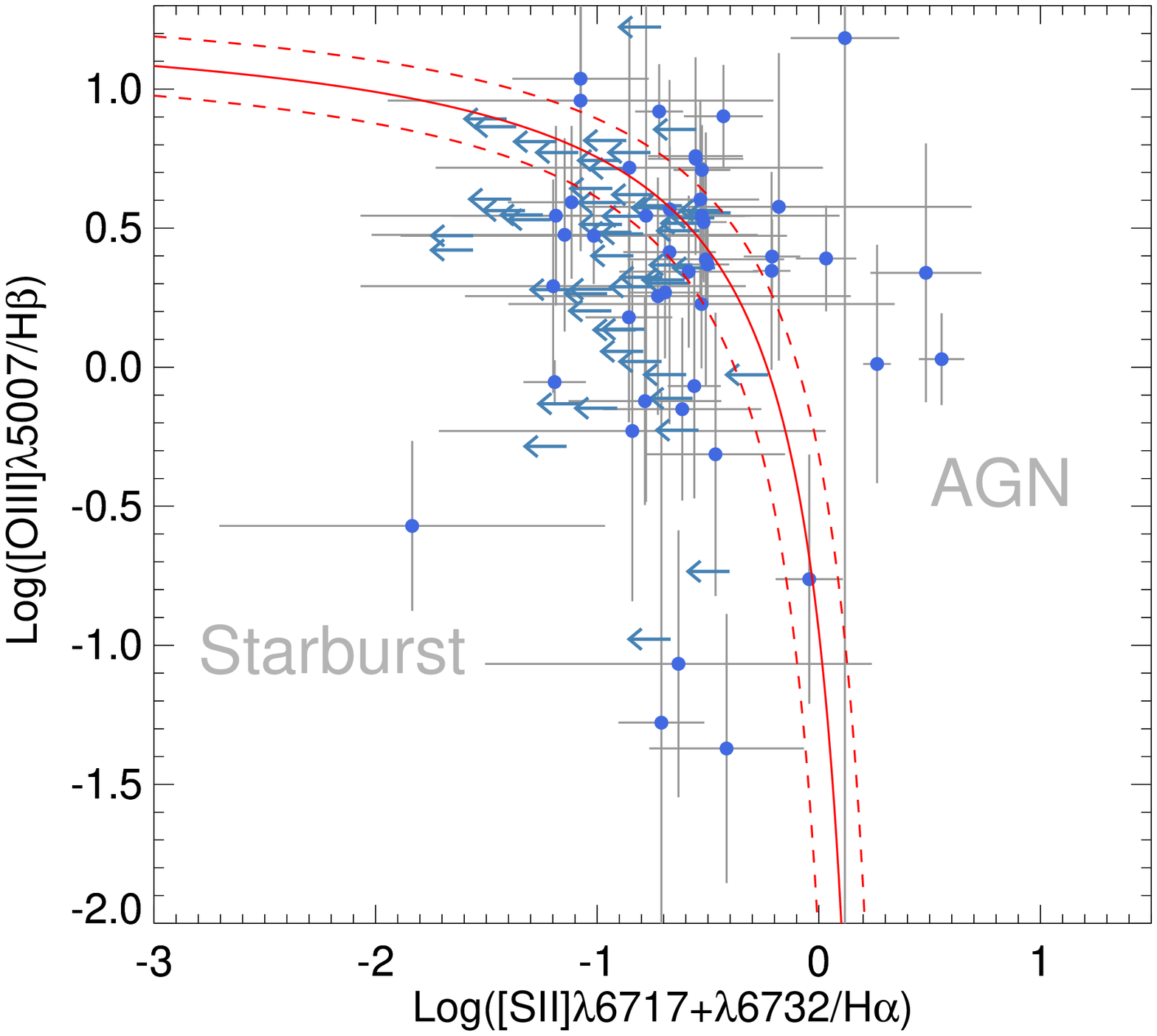} 	
   \includegraphics[width=8.9cm]{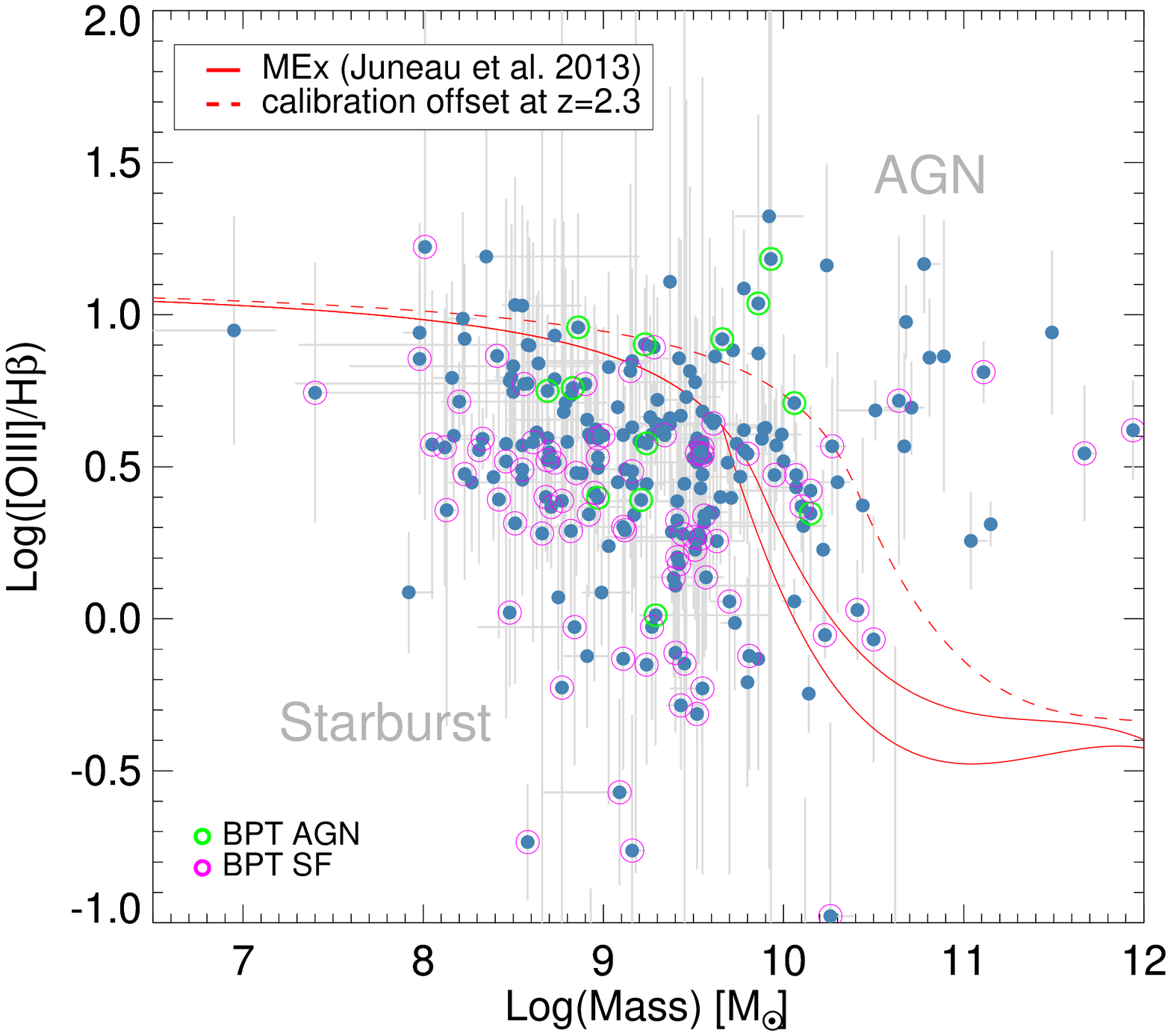} 
   \caption{AGN Diagnostics among star-forming galaxies. The {\it left panel} shows the BPT \citep{baldwin81} diagram for a subsample of galaxies for which [\sii]$\lambda$6717+$\lambda$6732/\ha\ and [\oiii]$\lambda$5007/\hb\ ratios are available (blue points) and for which the [\sii] line was not detected (blue arrows represent upper limits). The solid red curve represents the SF/AGN separation of \citet{kewley01}, while the dashed lines are $1-\sigma$ errors. We present in the {\it right panel} the Mass Excitation (MEx) diagram \citep[red solid curve,][]{juneau11}. This is similar to a BPT diagnostic except that the stellar mass is used as a proxy for the commonly used [\nii]/\ha\ ratio which is not available in our data. The dashed line is the new version of Juneau et al. relation scaled to $z \sim 2.3$. The green (magenta) circles are AGNs (SF galaxies) identified on the BPT diagram.}
   \vspace{1cm}
   \label{fig:agn}
\end{figure*}

\begin{table}
\caption{\label{tab:sample} HST Grism Sample}
\begin{tabular}{lccc}
 Field & \# Pointings & Original Catalog & Final Sample\\ \hline 
 WISPS & 29  & 1247 & 457  \\
 AEGIS & 28  & 405 & 95 \\
 COSMOS & 28  & 213 & 132  \\
 GOODS-S & 35  & 366 & 199  \\
 UDS & 28  & 226 & 151  \\ 
\vspace{0.3cm}
\end{tabular}
\end{table}

\subsection{[\nii] contamination}

The spectral resolution of the IR grism is too low to separate the [\nii]$\lambda$6548+6583 \AA\ lines from \ha, which may increase the observed \ha\ flux. Moreover, the [\nii] contamination appears to be increasing with stellar mass and with redshift \citep{erb06,sobral12,masters14}. Following \citet{dominguez13}, we estimate the [\nii] correction relative to \ha\ in bins of stellar mass by taking the average value of two approaches: {\it (i)} \citet{erb06} give an estimate of the [\nii] contribution at $z \sim 2$ in bins of stellar mass; {\it (ii)} \citet{sobral12} derive an empirical relation between EW$_{rest}$([\nii]+\ha) and [\nii]/\ha\ ratio using a large sample of SDSS galaxies. We divided our sample in three mass bins of log(\mstar/\msol) = 8, 9, and 10, for which the first method gives an [\nii] fraction of 0, 5, and 10\%. The second method yields a median [\nii] fraction of 12, 15, 23\%, respectively. Therefore, we adopted an [\nii] correction of 6, 10, and 16\% to correct our \ha\ fluxes in the three respective stellar mass bins. In Section \ref{sec:agn} we will see that in the determination of the AGN contribution the uncertainties from the [\nii] contamination are still smaller than those of the [\sii] or \hb\ line flux measurements.

\section{Sample Selection}
\label{sec:selection}

The galaxy sample resulting from the emission line selection in the two surveys described above contains 2457 galaxies. We select only galaxies with secure redshifts (quality flag Q1 and Q2) and free from contamination. Then we cross-match our redshift measurement with photometric and spectroscopic redshift information available from public catalogs. We used the NMBS v5.1 photometric catalog \citep{whitaker11} and the DEEP2 DR4 spectroscopic redshift catalog \citep{newman12} for the AEGIS field; the NMBS v5.1 for COSMOS; the FIREWORKS catalog \citep{wuyts08}, the CDFS photometric and spectro-$z$ compilation of N. Hathi (private communication) for GOODS-S; the photometric catalog of \citet{williams11}, spectro-$z$ and photo-$z$ catalogs of UKIDSS \citep{lawrence07} and CANDELS \citep{galametz13} for the UDS. All the galaxies that have Q1 flags in both catalogs are in good agreement. Galaxies with Q2 in our grism catalog are kept only if they are confirmed by spectro- or photo-$z$ information from other catalogs. After this quality selection, our final sample consists of 1034 galaxies. Table \ref{tab:sample} summarizes the sample size in each field before and after our selection procedure.

\section{Stellar Population Properties}
\label{sec:sed}

We used the catalogs described above for the SED fitting of the 3DHST galaxies. The WISP photometry consists of two IR bands (F110W, F160W), two UVIS bands (F475X, F600LP) and IRAC 3.6\micron\ band. The construction of the photometric catalog used here is described in \citet{bedregal13} and  \citet{colbert13}. Before we proceed to the SED fitting, we need to account for the contribution of nebular emission lines to the broadband fluxes. Several studies modeled or accounted for the emission line contribution, which can significantly change the stellar population properties inferred from population synthesis modeling \citep[e.g.][]{schaerer09, schaerer11, ono10, watson10, taniguchi10, mclure11, inoue11, finkelstein11,pacifici12}. In \citet{atek11}, we empirically demonstrated that the nebular contribution of galaxies having EW$_{rest} > 200$ \AA\ can lead to a brightening of 0.3 mag on average and up to 1 mag of their broadband flux. This can alter the estimate of stellar mass and age of galaxies by a factor of 2 on average and up to a factor of 10 \citep[see also][]{schenker13}. 

Following the same procedure presented in \citet{atek11}, we performed synthetic photometry with and without emission lines to estimate the nebular contribution in each galaxy spectrum. For the emission lines outside the spectral coverage of the grism, we used the typical emission line ratios observed in our data to infer the flux of those lines. We used a flux ratio of \ha/[OIII]$\lambda$5007=0.9 and [OIII]$\lambda$5007/[OII]$\lambda$3727=2.4. Then their contribution to the broadband flux density is estimated following the equations of \citet{guaita10} and \citet{finkelstein11}:

\begin{align} 
   f_{\lambda, em}={}& \frac{R_{T} \times f_{em}}{\int T_{\lambda} d\lambda}
\end{align}

where $R_{T}$ is the filter transmission at the emission line wavelength normalized by the maximum transmission of the filter, $f_{em}$ is the line flux inferred from the line ratio, and T$_{\lambda}$ is transmission curve of the filter. Finally, all the magnitudes in the corresponding filters are corrected in the different photometric catalogs.

The SED fitting procedure is performed using the {\rm FAST} code \citep{kriek09} with \citet{bc03} stellar population models. We used a \citet{chabrier03} initial mass function (IMF), a metallicity of $Z = 0.004$, a stellar attenuation in the range A$_{V}$=0-3 in steps of 0.1, and log(age/yr)= 7 - 10 in log steps of 0.1. An exponentially declining star formation model of the form exp($-t/\tau$) is assumed with log($\tau$/yr)=7-10 in steps of 0.2. The confidence levels of the resulting stellar properties are estimated by running a thousand Monte Carlo simulations.

\section{AGN Contribution}
\label{sec:agn}
The wavelength range covered by the WFC3 grisms is wide enough to simultaneously cover and resolve the optical lines used in the BPT diagnostic \citep{baldwin81} of AGN/SF separation only over a restricted redshift range, $0.8 \lesssim z \lesssim 1.4$ for WISP and around $z \sim 1.4$  ($\Delta z \sim 0.1$) for 3DHST.

The left panel of Fig. \ref{fig:agn} shows the [\oiii]$\lambda$5007/\hb\ ratio versus [\sii]$\lambda$6717+$\lambda$6732/\ha, for galaxies that have all the lines needed in the BPT diagram. The blue arrows represent upper limits in the case where the [\sii] line was not detected. The red solid line is the separation inferred from photoionization models by \citet{kewley01}. The red dashed line shows the 0.1 dex model uncertainties. We identify about 14\% of galaxies that appear to host an AGN. However, the AGN contamination is likely lower in our sample because of the redshift and mass dependance of the classification line. \citet{kewley13} proposed a revised version of the diagnostic line ratios that takes into account the redshift-evolution of the ISM conditions and ionizing radiation field up to $z \sim 2.5$. This is particularly relevant for our starburst sample were the extreme star formation may be seen as AGN activity according to the local calibration of the optical line ratios diagnostic. The correction shifts the classification line towards higher line ratios, hence decreasing the AGN fraction in our high-redshift sample. However, the update was for the [\nii]/\ha\ version of the classification, which cannot be quantified here.

The spectral resolution of the grisms is too low to resolve [\nii] and \ha, hence we use the Mass Excitation diagram presented in \citet{juneau11}, which uses the stellar mass as a proxy for the [\nii]/\ha\ ratio. This is mainly justified by the fact that [\nii]/\ha\ traces the gas phase metallicity and the strong correlation between the stellar mass and metallicity observed in star-forming galaxies \citep[e.g.][]{tremonti04,savaglio05,erb06}. Here we use the revised version of the MEx diagnostic (Juneau et al. 2013, private communication). To account for the fact that the MEx relation has been calibrated using local galaxies and for the redshift evolution of the mass-metallicity relation, Juneau et al. (2013) derived mass offsets as a function of redshift. The MEx diagram is shown in the right panel of Fig. \ref{fig:agn}, where only galaxies at $z \gtrsim 0.8$ (where \hb\ enters our spectral range in WISP data) are shown, together with the SF/AGN separation curve (solid line). The scaled MEx at $z \sim 2.3$ is shown by a dashed line.

We find that about 10\% of galaxies present ionizing characteristics of AGN. This fraction is comparable to the value derived from the BPT diagram. However, as illustrated by the green/magenta circles on the same figure, objects identified as AGN/SF in the BPT diagram can scatter to both sides of the AGN/SF demarcation line in the MEx diagram, which might be partially due to the large uncertainties in the [\oiii]$\lambda$5007/\hb\ and [\sii]$\lambda$6717+$\lambda$6732/\ha\ ratios observed in both figures, but also in the extrapolation of the MEx line from local galaxies to high-redshift. As stressed earlier, the AGN fraction derived from the BPT diagram should be considered as an upper limit, since we don not account for the redshift-evolution of the demarcation of \citet{kewley13}.

Using high resolution spectra of 22 star-forming galaxies at $z \sim$ 1.5-2, \citet{masters14} find no significant AGN activity based on BPT diagrams, although their galaxies have similar physical properties (SFR, stellar masses, etc) to our sample. While they observe an offset towards higher [\oiii]/\hb\ at a given [\nii]/\ha\ compared to local galaxies, it does not appear in the [\oiii]/\hb\ versus [\sii]/\ha\ diagram. They conclude that a high nitrogen abundance in high-redshift galaxies may explain such offset in the [\oiii]/\hb. The spectroscopic observations used in \citet{masters14} are certainly more reliable for AGN identification than our data, supporting the idea that our AGN fraction derived from the BPT diagram could be overestimated.

Additionally, we match the 3DHST fields with available X-ray source catalogs of the {\em Chandra} COSMOS survey \citep{elvis09} for COSMOS; AEGIS-X survey \citet{laird09} for AEGIS; the {\em Chandra} Deep Field South (CDFS) survey \citep{xue11} for GOODS-S; and SUBARU/XMM-NEWTON Deep Survey \citep[SXDS][]{ueda08} for UDS. We find that less than 1\% of galaxies show X-ray counterparts, with no source in common with the MEx or BPT diagnostics. X-ray observations are in general less sensitive to AGNs in low-mass galaxies, and the lack of detection does not systematically point to an absence of AGN. In the following, we excluded AGNs identified by the BPT diagram or X-ray observations.

\begin{figure}[!htbp]
   \centering
       \includegraphics[width=8.5cm]{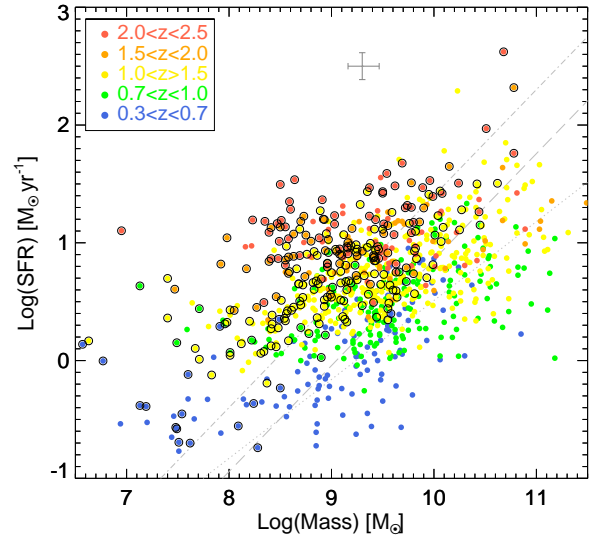}
   \caption{The SFR-mass relation for the emission-line galaxies up to the peak of star formation history. Data points are color-coded according to five redshift bins given in the legend. The lines correspond to the relation derived for $z < 0.7$ galaxies \citep[dotted line N07, ][]{noeske07}, $z \sim 1$ \citep[dashed line E07, ][]{elbaz07}, and $z \sim 2$\citep[dot-dashed D07, ][]{daddi07}. The black circles mark galaxies with very high equivalent width emission lines, EW$_{rest}$(\ha) $> 200$ \AA. The cross next to the legend presents the characteristic uncertainties of the sample in SFR and stellar mass.}
   \label{fig:sfr_mass_z}
\end{figure}

\begin{figure*}[!htbp]
   \centering
   \includegraphics[width=18cm]{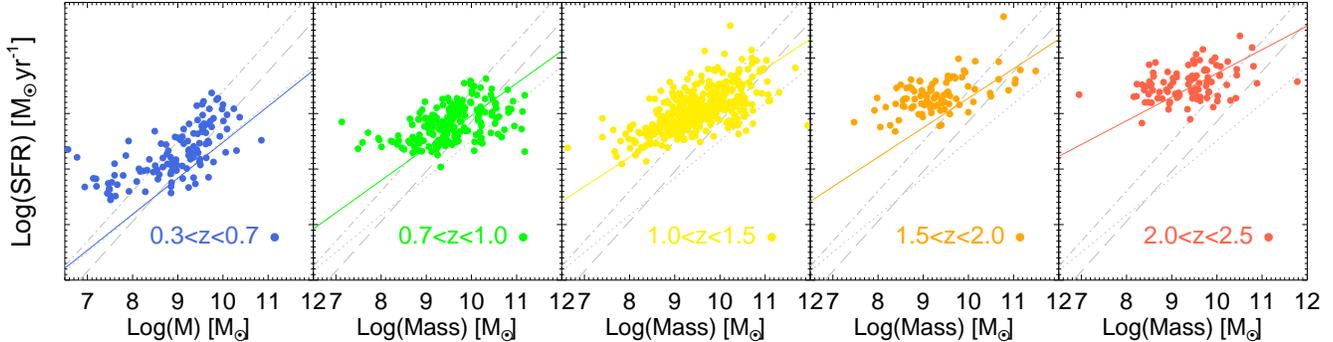} 
   \caption{Same as Fig. \ref{fig:sfr_mass_z} with SFR corrected for dust attenuation. The lines of N07, E07 and D07 are also shown in grey, while the colored lines following the same color code as the redshift bins are drawn from \citet{whitaker12}.}
   \label{fig:sfr_mass_cor}
\end{figure*}

\section{The Star Formation Mass Sequence}
\label{sec:sfr_mass}

The correlation between the star formation rate and the stellar mass of galaxies has been extensively studied over the past decade \citep[e.g.][]{noeske07,daddi07,elbaz07,rodighiero10,wuyts11,whitaker12}. However, the validity of the so-called ``main sequence'' (MS) of galaxies is still discussed, and so is the dispersion and the redshift-evolution of this correlation. Most of the studies at high-$z$ are based on continuum-selected samples that impose a stellar mass limit around $10^{10}$ \msol\ \citep{rodighiero11,whitaker12, salmi12}. The galaxy sample we use in the present study is emission-line selected, almost independently from the continuum brightness. This offers the possibility of studying the evolution of the SFR-Mass relation over a continuous redshift range from $0.3 \lesssim z \lesssim 2.4$ and down to a mass limit of 10$^{8}$ \msol. Most of the constraints on the SFR-M$_{\star}$ evolution previously reported at high-$z$ have been derived from continuum selected samples, where the SFR determination is based on the UV emission combined with the IR to account for both unobscured and obscured star formation. This approach exploits the radiation from a population of massive stars to estimate the averaged SFR over the past SF activity of the galaxy. We here use the nebular emission lines, which are the result of the photoionization of the neutral gas by young stars, to calculate the current SFR of galaxies using \citet{kennicutt98} calibration for the \ha\ line:

\begin{align}
SFR_{H\alpha} (M_{\odot} ~ yr^{-1}) = {}& 7.9 \times 10^{-42} L_{H\alpha} (erg ~ s^{-1})
 \end{align}
 
 In the case of 185 galaxies at $z \gtrsim 1.5$ that were selected by their strong [\oiii] emission, the \ha\ line was outside the wavelength coverage of the grism. Therefore, we use other emission lines to compute their SFR. For 127 of these galaxies the SFR was computed from the \hb\ line, assuming an \ha/\hb\ ratio of 2.86 \citep{osterbrock89} and applying the extinction correction derived later on in this section. For 20 of these galaxies the SFR was calculated using the [\oii]$\lambda$3727 line, adopting \citet{kennicutt98} calibration. As for the rest of galaxies at $z \gtrsim 1.5$ we used the [\oiii] line and a median ratio of \ha/[\oiii]$\lambda5007 \sim 0.9$ derived from our sample, with a standard deviation of 0.4. The conversion factor between the emission line luminosity and the SFR derived in \citet{kennicutt98} is a median value and the correct value can vary by 0.4 dex depending on the stellar mass. For each of our stellar mass bins we use a conversion factor of $3.6 \times 10^{-42}$, $4.2 \times 10^{-42}$, and $4.7 \times 10^{-42}$ from the low to the high mass bin, respectively, that were derived in \citet{brinchmann04}. To correct for the \citet{kroupa01} IMF adopted in \citet{brinchmann04}, we multiply these factors by 1.5 to be consistent with a \citet{salpeter55} IMF used in this work.

In Figure \ref{fig:sfr_mass_z}, we plot our sample of emission-line galaxies in the observed SFR-Mass plane, where the SFR is the observed value. The color-code for five redshift bins is defined in the legend. We also overplot the main MS relations derived at similar redshift bins by \citet[][in blue, at $z \sim 0.7$]{noeske07}, \citet[][in yellow, at $z \sim 1$]{elbaz07}, and \citet[][in red, at $z \sim 2$]{daddi07}. 
From the figure, it is clear that the emission-line selected galaxies are generally offset from each of the MS lines, i.e. have a higher SFR at a given stellar mass. We note that the derived relations in the literature account for the dust obscuration, whereas the SFR values presented in Fig. \ref{fig:sfr_mass_z} are not corrected for dust attenuation. We observe a clear redshift dependency as the normalization of the SFR-mass relation is increasing with $z$, from the blue to red sequence. Moreover, this evolution is observed over the same mass range of 8 $<$ log(\mstar/\msol) $<$ 11, confirming the previous results of mass-limited samples. 

In order to estimate the effects of dust on our results, we now proceed to the correction of the \ha\ emission for extinction. Because the Balmer ratio \ha/\hb\ is not accessible for all the galaxies, we rely on the correlation between the stellar mass and the dust content \citep[e.g.][]{pannella09,reddy10,bauer11,whitaker12} to estimate the extinction in our sample. We calculated the mean extinction from the \ha/\hb\ ratio for a subsample of 106 galaxies in bins of stellar mass [$<8, 8-9, 9-10, >10$] and found E(B-V) values of [0.05, 0.1, 0.18, 0.26], which we applied in the correction of all the galaxies in each of these bins. This result is consistent with the values derived by \citet{dominguez13} using the Balmer decrement in similar mass bins at $0.75 < z < 1.5$ or \citet{momcheva13} at $z \sim 0.8$. The extinction-corrected SFR-M relation is presented in Fig. \ref{fig:sfr_mass_cor} with the same color code as before. The main effect of the correction is to increase the specific star formation rate (sSFR = SFR/M$_{\star}$) of galaxies, and hence their offset from the MS. It also tends to steepen the slope of the SFR-M relation because the correction is more important for more massive galaxies. The log(SFR)-log(\mstar) equation has a slope of 0.65, shallower than a slope of 1 reported by \citet{wuyts11} at $z \sim 0-2.5$ or (0.77,0.9) by \citet{elbaz07} at $ z \sim$ (0,1) respectively. As shown earlier, the emission line selection introduces a lower limit on the SFR which may affect the SFR-\mstar\ slope, since low-mass and low-SFR galaxies may lie below the detection limit. The lower limit on SFR is typically between 0.3 and 10 \msolyr at $z \sim 0.5$ to 2.2.

We have plotted the recent results of \citet[][W12]{whitaker12} on the same figure, where each line has been derived at the mean value of each redshift bin. The star-forming sample of W12 was color-selected and the SFR was measured from the UV+IR emission. While the starburst galaxies are still offset from the main sequence, the W12 slope, which depends on the redshift following $0.7 - 0.13z$, is closer to our estimate, compared to the rest of the literature.

\begin{figure*}[htbp]
   \centering
       \includegraphics[width=18.5cm,angle=-90]{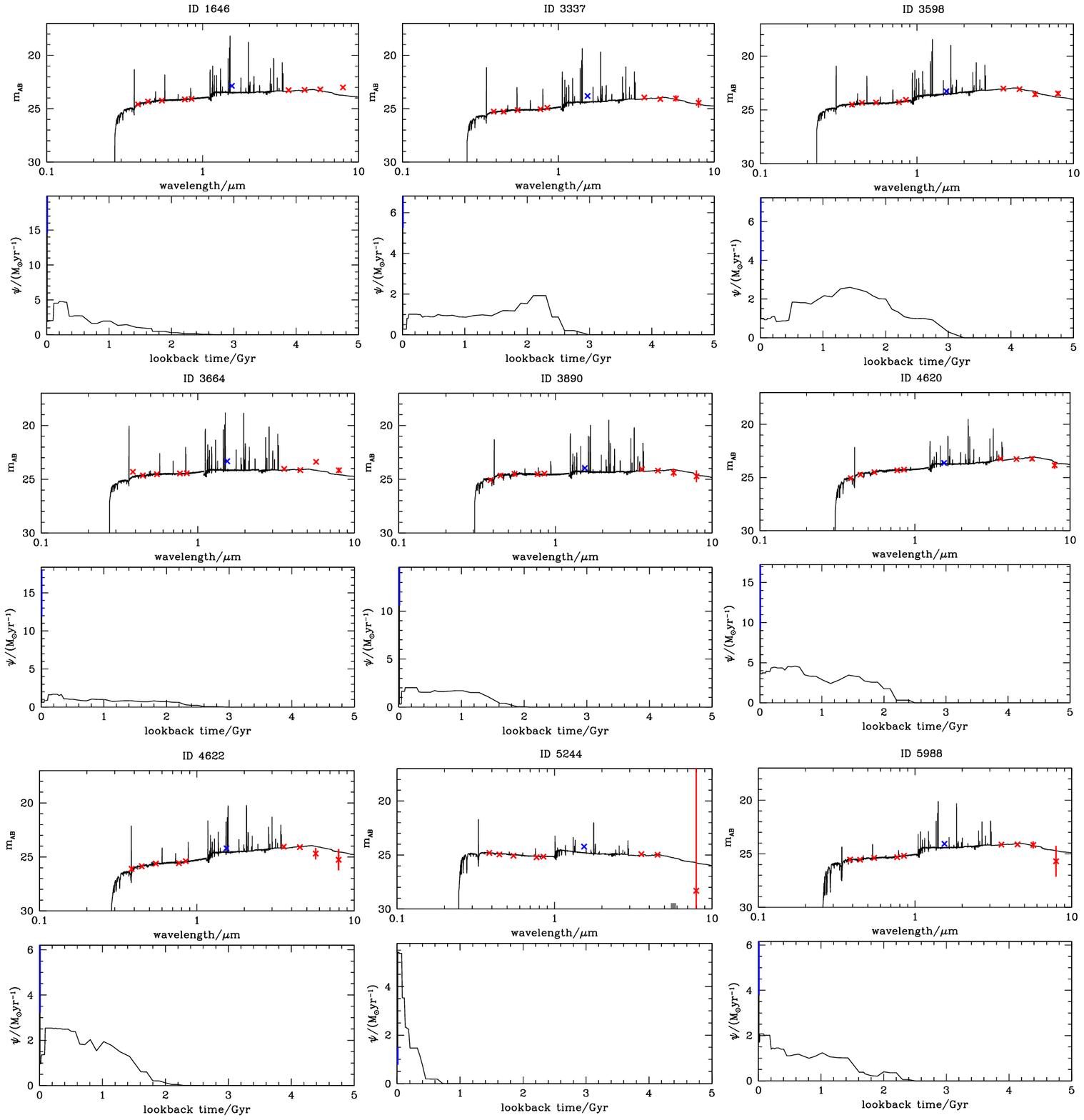}
   \caption{Star formation histories of starburst galaxies. For each object, we fit stellar population models of \citet{pacifici13} to the broadband photometry (with no correction for emission lines). {\it Top panel} shows for each galaxy the spectral energy distribution model (black curves) including emission lines and the observed magnitude in each band (red points and blue point for the {\em HST} H band). {\it Bottom panel} shows the SFR as a function of lookback time (where $t=0$ is set at $z=z_{gal}$) derived from our SFH models. The peak at the end of each curve is the current starburst constrained from the observed SFR based on emission lines, with the blue bar showing the associated uncertainties (16\% to 84\% interval in the probability distribution).}
   \label{fig:sed}
\end{figure*}

\section{Extreme Emission-Line Galaxies}
\label{sec:eelg}
In general, we show that by including the low-mass starburst galaxies, which were not easily accessible in previous observations, the dispersion of the SFR-M$\star$\ relation becomes more important. Galaxies with high-EW emission lines ($EW_{rest} > 200$ \AA) are marked with black circles in Figure \ref{fig:sfr_mass_z}. As we have shown in \citet{atek11}, this selection favors galaxies with high sSFR. Most of these galaxies are located on the high-SFR end of the figure, which would be considered as outliers to the MS relations of the literature. Extreme objects such as Ultra-Luminous IR Galaxies (ULIRG) and Submillimeter Galaxies (SMG) have been similarly identified as outliers in, for example, \citet{daddi10}, but are two orders of magnitude more massive than our low-mass starbursts. These EELGs are likely following different SF histories than the MS galaxies, as they experience intense starburst episodes over a short period of time. They can typically double their stellar mass in $\sim 100$ Myr or less. This high SF efficiency could be a direct consequence of the high gas fraction contained in these low-mass galaxies. Indeed, \citet{lara-lopez13a} showed the existence of a fundamental plane (FP) where the mass-metallicity relation evolves with SFR, and also with the \hi\ gas content, in the sense that high sSFR galaxies are also gas rich \citep[see also ][]{bothwell13}. They conclude that low-mass galaxies have a larger reservoir of \hi\ that fuels star formation over longer time scales leading to a slower enrichment of the ISM, compared to their massive counterparts. To place our results into this context, we will now investigate the star formation history of these extreme starbursts.

\subsection{Star Formation Histories of EELGs}

The irregular and stochastic SFHs expected in these starburst galaxies are generally not well approximated by a simple exponentially declining SFR \citep{lees09}. We use the spectral modeling approach of \citet{pacifici12} to constrain the spectral energy distributions and SFHs of such EELGs. The library is based on physically motivated star formation and chemical enrichment histories derived by performing a post-treatment of the Millennium cosmological simulation \citep{springel05} using the semi-analytic models of \citet{delucia07}. Similarly to \citet{pacifici13}, we build a library of one million model galaxies spanning observed redshfits in the range $0.5<z<2.5$, evolutionary stages up to $z=3$ \citep[see Sections 2.1 and 3.1.2 in][]{pacifici12}, current (averaged over the last 10 Myr) sSFR between 0.01 and 100 Gyr$^{-1}$, and current gas-phase oxygen abundance in the range $7 < 12$ + log (O/H) $< 9.4$. We then generate a library of model galaxy SEDs by combining this library of star formation and chemical enrichment histories with the latest version of the \citet{bc03} stellar population synthesis models, the nebular emission model of \citet{charlot01}, based on the photoionization code CLOUDY \citep{ferland96}, and the two-component dust model \`a la \citet{charlot00}. We adopt a Bayesian approach as in \citet{pacifici12} to extract best-estimate SFHs by comparing the SEDs of our model galaxies with the emission-line and broadband fluxes of 10 high-sSFR galaxies selected in the GOODS-S field to explore the stellar mass range of the sample. Specifically, we compare U, B, R, i, z, 3.6\micron, 4.5\micron, 5.8\micron, and 8\micron\ observed fluxes, with the fluxes of the model galaxies that lie within $\Delta z \sim 0.05$ of the spectroscopic redshift. In Fig. \ref{fig:sed}, we show for each galaxy of the subsample the best SED model (including nebular lines) compared to the observed fluxes (i.e. with no correction for emission line contribution), and the best-estimate SFR as a function of lookback time. For each galaxy, the reference $t = 0$ is fixed at its redshift $z_{gal}$. For the sake of consistency, we have compared the stellar masses derived from this sophisticated approach with the values obtained from the FAST fitting code of Sect. \ref{sec:sed}. The results are in good agreement and a significant deviation (more than 2$\sigma$) is seen for only two objects. The difference is due to an underestimate of the emission-line contribution in our SED modeling.

\begin{figure}[!htbp]
   \centering
       \includegraphics[width=9cm]{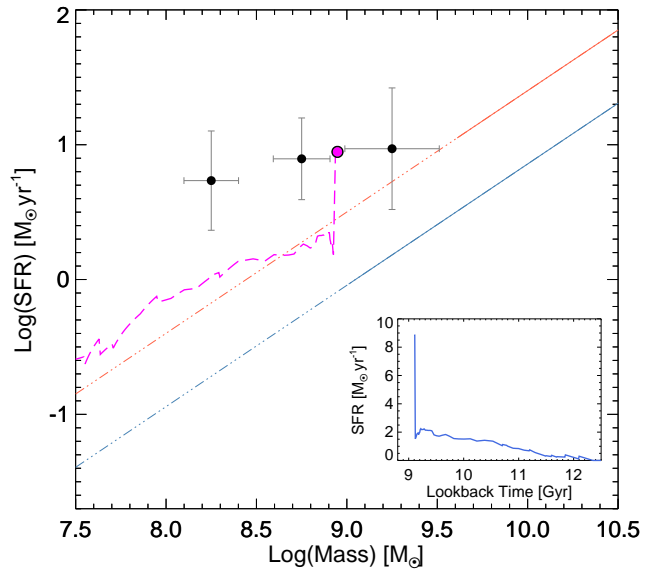}
   \caption{Starburst galaxies along the SFR-\mstar\ sequence. The magenta line shows the average SFR-\mstar\ evolution of a sample of high-sSFR galaxies. The black points are the average SFR in three mass bins of the total sample of emission-line galaxies over the same redshift range ($z > 1.5$) as the modeled galaxies. For reference, we show the main sequence derived at $z=1$ and 2 by \citet{elbaz07} and \citet{daddi07} with blue and red lines respectively. The inset shows the average SFH of our subsample of high-sSFR galaxies used to derive the SFR-\mstar\ evolution with time.}
   \label{fig:sfr_mass_sfh}
\end{figure}

In the inset of Fig. \ref{fig:sfr_mass_sfh}, we show the averaged SFH of this sub-sample of high-sSFR galaxies with redshifts in the range $1.5 < z < 2.3$ and stellar masses between $10^{8.5} < M_{\star} < 10^{9.5}$ \msol. The SFR is plotted against the lookback time, where $t=0$ is set at $z_{gal}$. We see a gradually rising SFR from the formation epoch while galaxies show an ongoing starburst episode on top of the longer time-scale SFR. Although the SFH of individual galaxies may have large uncertainties, the averaged SFH is more likely to be representative of this class of low-mass galaxies. While the current burst of each galaxy is constrained by the \ha\ emission, the time resolution in the SF models is not sufficient to isolate the previous bursts which are smoothed over the past SF activity. Therefore, it is possible that the average increase of SF with time is in fact an increase in the amplitude of successive short bursts. However, our semi-analytical treatment of cosmological simulations rarely predicts short-lived strong bursts but rather SF fluctuations arising from gas infall, feedback and merger histories.  Only the last burst seen on top of the rising SFH is observationally constrained from the \ha\ line. The investigation of the model fits show that the typical duration of the burst is between 10 and 100 Myr and the last SFR peak at $z \sim 1.5$, for example, is five times higher than the average SFR.

Similarly, using high-resolution cosmological simulations of dwarf galaxies, \citet{shen13} show that they have an extended and stochastic star-formation process with strong and short-lived bursts. During the starburst period, the sSFR peaks at 50-100 Gyr$^{-1}$, in agreement with the strongest bursts in our EELG sample. Each starburst phase is preceded by an increase in the density of cold gas accreted from the halo, which triggers star formation. The interplay between gas accretion and supernovae-induced outflows is at the origin of the stochastic SF events. Among the mechanisms that can contribute to the onset of such bursts, galaxy interactions and mergers are known to play an important role in gas infalls \citep{dimatteo08}. \citet{hung13} also show that high-sSFR galaxies include a larger number of interacting and disturbed systems compared to normal SF galaxies, supporting the important role of merger-induced starbursts.

We now use these SFH constraints to follow the evolution of EELGs along the SFR-\mstar\ sequence. Figure \ref{fig:sfr_mass_sfh} presents the average evolution (magenta curve) of our starburst subsample in this plane until the redshift at which it is observed, compared with the rest of the sample at comparable redshifts ($z  > 1.5$). It shows the average stellar mass buildup with time of starburst galaxies. We can see how galaxies stay close to the MS with a similar slope until a burst event increases the SFR to values well above the level of normal SF population. The modeled SFH is in good agreement with the observed average SFR in three low-mass bins of all emission-line selected galaxies at $z > 1.5$ represented by the black points. As we pointed out earlier, we are able to constrain only the last burst of SF, which moves the galaxy outside the MS. The succession of short starbursts can make galaxies bounce in and out of the MS rather than a smooth evolution along it, because of their short dynamical time scales and efficient outflows. While massive galaxies are characterized by secular star formation adequately described by the SF main sequence, low-mass galaxies experience large excursions outside the MS due to stochastic and powerful star formation events.

\begin{figure}[htbp]
   \centering
      \includegraphics[width=9cm]{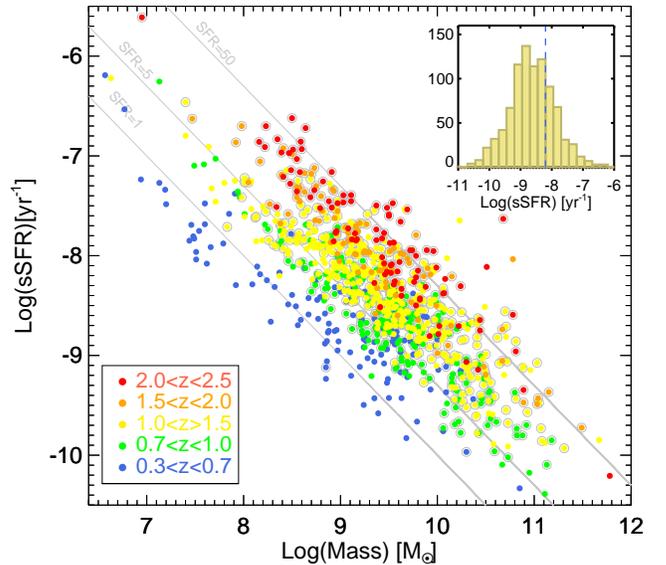}
   \caption{The specific star formation rate as a function of the stellar mass. The color code for the redshift bins is the same as in Fig. \ref{fig:sfr_mass_cor}. The solid lines represent constant star formation of 1, 5, and 50 \msolyr. The grey circles denote galaxies from the 3DHST sample. The inset shows the distribution of the sSFR for the whole sample where the dashed vertical line marks the threshold adopted to identify starburst galaxies.}
   \label{fig:ssfr_mass}
\end{figure}

\subsection{The Role of Starbursts in the Star Formation History at High Redshift}

To assess the relative importance of the starburst galaxies, we derived the specific SFR for our full sample of emission-line galaxies, which is presented as a function of stellar mass in Fig. \ref{fig:ssfr_mass}. Given the shallow slope of the SFR-\mstar\ relation, the galaxies nearly follow the constant SFR line (solid lines) on the sSFR-\mstar\ plane. We have marked with grey circles the 3DHST sample of galaxies. As expected, because of the difference in the spectral coverage between WISP and 3DHST, the circles do not cover the lower redshift bin (in blue). The classification of starbursts with respect to the MS star-forming galaxies cannot be easily established from the SFR-\mstar\ relation, since this emission-line selected sample defines, in each redshift bin, a distinct sequence offset from the MS. One widely used criterion to define a starburst is to apply an \ewha\ cut around 100 \AA\ or a birthrate parameter {\em b =SFR/$<SFR>$} higher than 2-3 \citep{scalo86,kennicutt98,lee09}.
Therefore, the \ha\ equivalent width threshold of 200 \AA, which we adopted earlier to identify outliers to the MS, ensures the selection of extreme starbursts and has the advantage of being purely observational, i.e. independent from any model assumptions. We note that the [\oiii] equivalent width threshold is higher according to the median \ha/[\oiii] ratio derived earlier.

Now, in order to compare our results to those of \citet{rodighiero11} who specifically estimated the contribution of starbursts to the total SFR density at $z \sim 2$, we choose a similar selection of starburst galaxies based on the specific SFR. The sSFR corresponds to a measure of the current SFR over the total stellar mass assembled during the past star formation of the galaxy. For comparison, applying their criterion of log(sSFR)$= -8.2$ to our sample results in an \ewha\ threshold of about 80 \AA, accounting for significant uncertainties on both the stellar mass and the [\oiii]/\ha\ ratio when \ha\ is not available. 

By selecting off-sequence galaxies, \citet{rodighiero11} have reported that the starburst population accounts for only 10\% of the total SFR density at $1.5 < z < 2.5$ in the high-mass regime at \mstar$> 10^{10}$\msol. For that study, they combined a FIR-selected sample (SFR-limited) with a BzK-selected \citep{daddi07} sample (mass-limited). Our emission-line selection has its own completeness limits, which tend to be complementary to other selection methods. 

More massive galaxies will have a brighter continuum, which reduces the contrast (equivalent width) of the line for a given line flux limit. This introduces an upper limit limit in stellar mass of about 10$^{10}$ \msol, deduced from the combination of the EW and line flux limits, accounting for dust attenuation at those stellar masses. Lower mass galaxies, although detected by their emission lines, are more likely to be missing in the photometric catalogs used for the SED fitting, putting a lower mass limit of $\sim 10^{8.2}$ \msol, mainly imposed by the depth of the imaging data. From the line flux limit the SFR incompleteness of our selection is about 2 \msolyr at $z = 1$ and 10 \msolyr at $z = 2$. Our data show that the contribution of starburst galaxies with \ewha\ larger than 300, 200, and 100 \AA\ to the total SF in the mass range $10^{8.2} <$ \mstar $< 10^{10}$ \msol\ at $z=1-2$ is around 13\%, 18\%, and 34\%, respectively. Although probing a different mass range, the sample of \citet{rodighiero11} shows an increase towards lower masses (down to 10$^{10}$ \msol) in the contribution of off-sequence galaxies to the total SFR density. 

In the inset of Fig. \ref{fig:ssfr_mass}, we show the sSFR distribution, where the dashed line denotes log(sSFR)= -8.2. At this SFR value, the galaxy will be able to double its stellar mass within $\sim 150$ Myr, which is much shorter than its typical age of 1-3 Gyrs we derived from SFH modeling (cf. Sect. \ref{sec:sfr_mass}). This is the typical value adopted in \citet{rodighiero11} to separate starburst and ``normal'' star-forming galaxies. Comparing the total SF produced by galaxies having log(sSFR/yr$^{-1}$) $> -8.2$  with the SF of the whole sample in the range of completeness described above, we find that starburst galaxies account for 29\% of the total SF.

 It is clear that in the stellar mass range below 10$^{10}$ \msol, the role of starburst galaxies is more important than what has been established for more massive galaxies. Moreover, the UV luminosity function at $z \sim 2$ shows no evidence of turnover at very faint UV magnitudes \citep[e.g.,][]{Alavi13}, supporting the idea that an important part of the star formation may occur in dwarf galaxies. We also observe that the prevalence of young, and presumably starbursting galaxies, increases with redshift as suggested by the increase in \ewha\ and the number density of EELGs \citep{atek11,shim11,fumagalli12}.

\section{Conclusion}
\label{sec:conclusion}
Using {\em HST} IR grism observations, we analyzed a large sample of extreme starburst galaxies at $0.3 < z < 2.3$. The WFC3 slitless spectroscopy enables us to extend the SFR-\mstar\ relation to low-mass galaxies by an order of magnitude compared to previous studies at high redshift. The emission-line selection tend to harvest more outliers to the SF ``main sequence'' than previous studies, which increases the dispersion and flattens the slope of the SF sequence.

We identify extreme emission-line galaxies that show equivalent width between 200 and 1500 \AA. This selection favors high-sSFR galaxies that, for a given mass, form stars at a much higher rate than normal star-forming galaxies. Using stellar population models we put constraints on the SF histories of these extreme starbursts. For most of these galaxies, the SFR rises continuously and follows the SFR-\mstar\ relation from their formation epoch ($\sim 2-3$ Gyr ago) until the most recent burst of star formation, which brings them above the main SF sequence. This may be caused by a succession of gas accretion and starburst episodes leading to a stochastic mode of SF in these low-mass galaxies. 

To assess the contribution of starburst galaxies to the star formation density we applied selection thresholds in the line equivalent width. We show that dwarf galaxies play an important role during the peak epoch of star formation history. At $1<z<2$, we find that 13\%, 18\%, and 34\% of the total star formation of the emission-line selected sample in the mass range $10^{8.2} < M_{\star} < 10^{10}$ is produced by galaxies having \ewha\ larger than 300 \AA, 200 \AA, and 100 \AA, respectively. For comparison, the starburst selection criterion used in \citet{rodighiero11}, i.e. log(sSFR) $= -8.2$, corresponds to mass-doubling time scale of $ \sim 150$ Myr and a EW cut of about 80 \AA. Applying this threshold to our sample results in a contribution of 29\% to the total SFR density.  

More star formation may occur in even lower-mass galaxies having higher \ewha\ and that dominate the UV luminosity function at $z \sim 2$ \citep{fumagalli12, Alavi13}. High-resolution NIR spectroscopic observations,with new instruments such as {\em VLT}/KMOS or {\em Keck}/MOSFIRE, will help us put better constrains on the SF properties and metallicities of this population. Deeper imaging will also be needed to constrain their stellar mass to assess the SF budget of lower mass galaxies at the peak of star formation history of the Universe.

\acknowledgments

HA and JPK are supported by the European Research Council (ERC) advanced grant ``Light on the Dark'' (LIDA). C.P. acknowledges support from the KASI-Yonsei Joint Research Program for the Frontiers of Astronomy and Space Science funded by the Korea Astronomy and Space Science Institute. SC acknowledges support from the European Research Council via an Advanced Grant under grant agreement no. 321323.

\bibliographystyle{apj}
\bibliography{references}

\begin{thebibliography}{90}
\expandafter\ifx\csname natexlab\endcsname\relax\def\natexlab#1{#1}\fi

\bibitem[{{Alavi} {et~al.}(2013){Alavi}, {Siana}, {Richard}, {Stark},
  {Scarlata}, {Teplitz}, {Freeman}, {Dominguez}, {Rafelski}, {Robertson}, \&
  {Kewley}}]{Alavi13}
{Alavi}, A., {et~al.} 2013, ArXiv e-prints

\bibitem[{{Amor{\'{\i}}n} {et~al.}(2010){Amor{\'{\i}}n}, {P{\'e}rez-Montero},
  \& {V{\'{\i}}lchez}}]{amorin10}
{Amor{\'{\i}}n}, R.~O., {P{\'e}rez-Montero}, E., \& {V{\'{\i}}lchez}, J.~M.
  2010, \apjl, 715, L128

\bibitem[{{Atek} {et~al.}(2010){Atek}, {Malkan}, {McCarthy}, {Teplitz},
  {Scarlata}, {Siana}, {Henry}, {Colbert}, {Ross}, {Bridge}, {Bunker},
  {Dressler}, {Fosbury}, {Martin}, \& {Shim}}]{atek10}
{Atek}, H., {et~al.} 2010, \apj, 723, 104

\bibitem[{{Atek} {et~al.}(2011){Atek}, {Siana}, {Scarlata}, {Malkan},
  {McCarthy}, {Teplitz}, {Henry}, {Colbert}, {Bridge}, {Bunker}, {Dressler},
  {Fosbury}, {Hathi}, {Martin}, {Ross}, \& {Shim}}]{atek11}
---. 2011, \apj, 743, 121

\bibitem[{{Baldwin} {et~al.}(1981){Baldwin}, {Phillips}, \&
  {Terlevich}}]{baldwin81}
{Baldwin}, J.~A., {Phillips}, M.~M., \& {Terlevich}, R. 1981, \pasp, 93, 5

\bibitem[{{Bauer} {et~al.}(2011){Bauer}, {Conselice}, {Perez-Gonzalez},
  {Grutzbauch}, {Bluck}, {Buitrago}, \& {Mortlock}}]{bauer11}
{Bauer}, A.~E., {Conselice}, C.~J., {Perez-Gonzalez}, P.~G., {Grutzbauch}, R.,
  {Bluck}, A.~F.~L., {Buitrago}, F., \& {Mortlock}, A. 2011, ArXiv e-prints

\bibitem[{{Bedregal} {et~al.}(2013){Bedregal}, {Scarlata}, {Henry}, {Atek},
  {Rafelski}, {Teplitz}, {Dominguez}, {Siana}, {Colbert}, {Malkan}, {Ross},
  {Martin}, {Dressler}, {Bridge}, {Hathi}, {Masters}, {McCarthy}, \&
  {Rutkowski}}]{bedregal13}
{Bedregal}, A.~G., {et~al.} 2013, \apj, 778, 126

\bibitem[{{Bothwell} {et~al.}(2013){Bothwell}, {Maiolino}, {Kennicutt},
  {Cresci}, {Mannucci}, {Marconi}, \& {Cicone}}]{bothwell13}
{Bothwell}, M.~S., {Maiolino}, R., {Kennicutt}, R., {Cresci}, G., {Mannucci},
  F., {Marconi}, A., \& {Cicone}, C. 2013, \mnras, 433, 1425

\bibitem[{{Bouch{\'e}} {et~al.}(2010){Bouch{\'e}}, {Dekel}, {Genzel}, {Genel},
  {Cresci}, {F{\"o}rster Schreiber}, {Shapiro}, {Davies}, \&
  {Tacconi}}]{bouche10}
{Bouch{\'e}}, N., {et~al.} 2010, \apj, 718, 1001

\bibitem[{{Brammer} {et~al.}(2012){Brammer}, {van Dokkum}, {Franx},
  {Fumagalli}, {Patel}, {Rix}, {Skelton}, {Kriek}, {Nelson}, {Schmidt},
  {Bezanson}, {da Cunha}, {Erb}, {Fan}, {F{\"o}rster Schreiber}, {Illingworth},
  {Labb{\'e}}, {Leja}, {Lundgren}, {Magee}, {Marchesini}, {McCarthy},
  {Momcheva}, {Muzzin}, {Quadri}, {Steidel}, {Tal}, {Wake}, {Whitaker}, \&
  {Williams}}]{brammer12}
{Brammer}, G.~B., {et~al.} 2012, \apjs, 200, 13

\bibitem[{{Brinchmann} {et~al.}(2004){Brinchmann}, {Charlot}, {White},
  {Tremonti}, {Kauffmann}, {Heckman}, \& {Brinkmann}}]{brinchmann04}
{Brinchmann}, J., {Charlot}, S., {White}, S.~D.~M., {Tremonti}, C.,
  {Kauffmann}, G., {Heckman}, T., \& {Brinkmann}, J. 2004, \mnras, 351, 1151

\bibitem[{{Bruzual} \& {Charlot}(2003)}]{bc03}
{Bruzual}, G., \& {Charlot}, S. 2003, \mnras, 344, 1000

\bibitem[{{Cardamone} {et~al.}(2009){Cardamone}, {Schawinski}, {Sarzi},
  {Bamford}, {Bennert}, {Urry}, {Lintott}, {Keel}, {Parejko}, {Nichol},
  {Thomas}, {Andreescu}, {Murray}, {Raddick}, {Slosar}, {Szalay}, \&
  {Vandenberg}}]{cardamone09}
{Cardamone}, C., {et~al.} 2009, \mnras, 399, 1191

\bibitem[{{Chabrier}(2003)}]{chabrier03}
{Chabrier}, G. 2003, \apjl, 586, L133

\bibitem[{{Charlot} \& {Fall}(2000)}]{charlot00}
{Charlot}, S., \& {Fall}, S.~M. 2000, \apj, 539, 718

\bibitem[{{Charlot} \& {Longhetti}(2001)}]{charlot01}
{Charlot}, S., \& {Longhetti}, M. 2001, \mnras, 323, 887

\bibitem[{{Colbert} {et~al.}(2013){Colbert}, {Teplitz}, {Atek}, {Bunker},
  {Rafelski}, {Ross}, {Scarlata}, {Bedregal}, {Dominguez}, {Dressler}, {Henry},
  {Malkan}, {Martin}, {Masters}, {McCarthy}, \& {Siana}}]{colbert13}
{Colbert}, J.~W., {et~al.} 2013, ArXiv e-prints

\bibitem[{{Daddi} {et~al.}(2007){Daddi}, {Dickinson}, {Morrison}, {Chary},
  {Cimatti}, {Elbaz}, {Frayer}, {Renzini}, {Pope}, {Alexander}, {Bauer},
  {Giavalisco}, {Huynh}, {Kurk}, \& {Mignoli}}]{daddi07}
{Daddi}, E., {et~al.} 2007, \apj, 670, 156

\bibitem[{{Daddi} {et~al.}(2010){Daddi}, {Elbaz}, {Walter}, {Bournaud},
  {Salmi}, {Carilli}, {Dannerbauer}, {Dickinson}, {Monaco}, \&
  {Riechers}}]{daddi10}
---. 2010, \apjl, 714, L118

\bibitem[{{De Lucia} \& {Blaizot}(2007)}]{delucia07}
{De Lucia}, G., \& {Blaizot}, J. 2007, \mnras, 375, 2

\bibitem[{{Dekel} {et~al.}(2009){Dekel}, {Birnboim}, {Engel}, {Freundlich},
  {Goerdt}, {Mumcuoglu}, {Neistein}, {Pichon}, {Teyssier}, \&
  {Zinger}}]{dekel09}
{Dekel}, A., {et~al.} 2009, \nat, 457, 451

\bibitem[{{Di Matteo} {et~al.}(2008){Di Matteo}, {Bournaud}, {Martig},
  {Combes}, {Melchior}, \& {Semelin}}]{dimatteo08}
{Di Matteo}, P., {Bournaud}, F., {Martig}, M., {Combes}, F., {Melchior}, A.-L.,
  \& {Semelin}, B. 2008, \aap, 492, 31

\bibitem[{{Dom{\'{\i}}nguez} {et~al.}(2013){Dom{\'{\i}}nguez}, {Siana},
  {Henry}, {Scarlata}, {Bedregal}, {Malkan}, {Atek}, {Ross}, {Colbert},
  {Teplitz}, {Rafelski}, {McCarthy}, {Bunker}, {Hathi}, {Dressler}, {Martin},
  \& {Masters}}]{dominguez13}
{Dom{\'{\i}}nguez}, A., {et~al.} 2013, \apj, 763, 145

\bibitem[{{Elbaz} {et~al.}(2007){Elbaz}, {Daddi}, {Le Borgne}, {Dickinson},
  {Alexander}, {Chary}, {Starck}, {Brandt}, {Kitzbichler}, {MacDonald},
  {Nonino}, {Popesso}, {Stern}, \& {Vanzella}}]{elbaz07}
{Elbaz}, D., {et~al.} 2007, \aap, 468, 33

\bibitem[{{Elvis} {et~al.}(2009){Elvis}, {Civano}, {Vignali}, {Puccetti},
  {Fiore}, {Cappelluti}, {Aldcroft}, {Fruscione}, {Zamorani}, {Comastri},
  {Brusa}, {Gilli}, {Miyaji}, {Damiani}, {Koekemoer}, {Finoguenov}, {Brunner},
  {Urry}, {Silverman}, {Mainieri}, {Hasinger}, {Griffiths}, {Carollo}, {Hao},
  {Guzzo}, {Blain}, {Calzetti}, {Carilli}, {Capak}, {Ettori}, {Fabbiano},
  {Impey}, {Lilly}, {Mobasher}, {Rich}, {Salvato}, {Sanders}, {Schinnerer},
  {Scoville}, {Shopbell}, {Taylor}, {Taniguchi}, \& {Volonteri}}]{elvis09}
{Elvis}, M., {et~al.} 2009, \apjs, 184, 158

\bibitem[{{Erb} {et~al.}(2006){Erb}, {Shapley}, {Pettini}, {Steidel}, {Reddy},
  \& {Adelberger}}]{erb06}
{Erb}, D.~K., {Shapley}, A.~E., {Pettini}, M., {Steidel}, C.~C., {Reddy},
  N.~A., \& {Adelberger}, K.~L. 2006, \apj, 644, 813

\bibitem[{{Ferland}(1996)}]{ferland96}
{Ferland}, G.~J. 1996, {Hazy, A Brief Introduction to Cloudy 90}

\bibitem[{{Finkelstein} {et~al.}(2011){Finkelstein}, {Hill}, {Gebhardt},
  {Adams}, {Blanc}, {Papovich}, {Ciardullo}, {Drory}, {Gawiser}, {Gronwall},
  {Schneider}, \& {Tran}}]{finkelstein11}
{Finkelstein}, S.~L., {et~al.} 2011, \apj, 729, 140

\bibitem[{{Fumagalli} {et~al.}(2012){Fumagalli}, {Patel}, {Franx}, {Brammer},
  {van Dokkum}, {da Cunha}, {Kriek}, {Lundgren}, {Momcheva}, {Rix}, {Schmidt},
  {Skelton}, {Whitaker}, {Labbe}, \& {Nelson}}]{fumagalli12}
{Fumagalli}, M., {et~al.} 2012, \apjl, 757, L22

\bibitem[{{Galametz} {et~al.}(2013){Galametz}, {Grazian}, {Fontana},
  {Ferguson}, {Ashby}, {Barro}, {Castellano}, {Dahlen}, {Donley}, {Faber},
  {Grogin}, {Guo}, {Huang}, {Kocevski}, {Koekemoer}, {Lee}, {McGrath}, {Peth},
  {Willner}, {Almaini}, {Cooper}, {Cooray}, {Conselice}, {Dickinson}, {Dunlop},
  {Fazio}, {Foucaud}, {Gardner}, {Giavalisco}, {Hathi}, {Hartley}, {Koo},
  {Lai}, {de Mello}, {McLure}, {Lucas}, {Paris}, {Pentericci}, {Santini},
  {Simpson}, {Sommariva}, {Targett}, {Weiner}, {Wuyts}, \& {the CANDELS
  Team}}]{galametz13}
{Galametz}, A., {et~al.} 2013, \apjs, 206, 10

\bibitem[{{Grogin} {et~al.}(2011){Grogin}, {Kocevski}, {Faber}, {Ferguson},
  {Koekemoer}, {Riess}, {Acquaviva}, {Alexander}, {Almaini}, {Ashby}, {Barden},
  {Bell}, {Bournaud}, {Brown}, {Caputi}, {Casertano}, {Cassata}, {Challis},
  {Chary}, {Cheung}, {Cirasuolo}, {Conselice}, {Roshan Cooray}, {Croton},
  {Daddi}, {Dahlen}, {Dav{\'e}}, {de Mello}, {Dekel}, {Dickinson}, {Dolch},
  {Donley}, {Dunlop}, {Dutton}, {Elbaz}, {Fazio}, {Filippenko}, {Finkelstein},
  {Fontana}, {Gardner}, {Garnavich}, {Gawiser}, {Giavalisco}, {Grazian}, {Guo},
  {Hathi}, {H{\"a}ussler}, {Hopkins}, {Huang}, {Huang}, {Jha}, {Kartaltepe},
  {Kirshner}, {Koo}, {Lai}, {Lee}, {Li}, {Lotz}, {Lucas}, {Madau}, {McCarthy},
  {McGrath}, {McIntosh}, {McLure}, {Mobasher}, {Moustakas}, {Mozena}, {Nandra},
  {Newman}, {Niemi}, {Noeske}, {Papovich}, {Pentericci}, {Pope}, {Primack},
  {Rajan}, {Ravindranath}, {Reddy}, {Renzini}, {Rix}, {Robaina}, {Rodney},
  {Rosario}, {Rosati}, {Salimbeni}, {Scarlata}, {Siana}, {Simard}, {Smidt},
  {Somerville}, {Spinrad}, {Straughn}, {Strolger}, {Telford}, {Teplitz},
  {Trump}, {van der Wel}, {Villforth}, {Wechsler}, {Weiner}, {Wiklind}, {Wild},
  {Wilson}, {Wuyts}, {Yan}, \& {Yun}}]{grogin11}
{Grogin}, N.~A., {et~al.} 2011, ArXiv e-prints

\bibitem[{{Guaita} {et~al.}(2010){Guaita}, {Gawiser}, {Padilla}, {Francke},
  {Bond}, {Gronwall}, {Ciardullo}, {Feldmeier}, {Sinawa}, {Blanc}, \&
  {Virani}}]{guaita10}
{Guaita}, L., {et~al.} 2010, \apj, 714, 255

\bibitem[{{Hathi} {et~al.}(2013){Hathi}, {Cohen}, {Ryan}, {Finkelstein},
  {McCarthy}, {Windhorst}, {Yan}, {Koekemoer}, {Rutkowski}, {O'Connell},
  {Straughn}, {Balick}, {Bond}, {Calzetti}, {Disney}, {Dopita}, {Frogel},
  {Hall}, {Holtzman}, {Kimble}, {Paresce}, {Saha}, {Silk}, {Trauger}, {Walker},
  {Whitmore}, \& {Young}}]{hathi13}
{Hathi}, N.~P., {et~al.} 2013, \apj, 765, 88

\bibitem[{{Henry} {et~al.}(2013){Henry}, {Scarlata}, {Dom{\'{\i}}nguez},
  {Malkan}, {Martin}, {Siana}, {Atek}, {Bedregal}, {Colbert}, {Rafelski},
  {Ross}, {Teplitz}, {Bunker}, {Dressler}, {Hathi}, {Masters}, {McCarthy}, \&
  {Straughn}}]{henry13b}
{Henry}, A., {et~al.} 2013, \apjl, 776, L27

\bibitem[{{Hopkins}(2004)}]{hopkins04}
{Hopkins}, A.~M. 2004, \apj, 615, 209

\bibitem[{{Hung} {et~al.}(2013){Hung}, {Sanders}, {Casey}, {Lee}, {Barnes},
  {Capak}, {Kartaltepe}, {Koss}, {Larson}, {Le Floc'h}, {Lockhart}, {Man},
  {Mann}, {Riguccini}, {Scoville}, \& {Symeonidis}}]{hung13}
{Hung}, C.-L., {et~al.} 2013, ArXiv e-prints

\bibitem[{{Inoue}(2011)}]{inoue11}
{Inoue}, A.~K. 2011, ArXiv e-prints

\bibitem[{{Juneau} {et~al.}(2011){Juneau}, {Dickinson}, {Alexander}, \&
  {Salim}}]{juneau11}
{Juneau}, S., {Dickinson}, M., {Alexander}, D.~M., \& {Salim}, S. 2011, \apj,
  736, 104

\bibitem[{{Kakazu} {et~al.}(2007){Kakazu}, {Cowie}, \& {Hu}}]{kakazu07}
{Kakazu}, Y., {Cowie}, L.~L., \& {Hu}, E.~M. 2007, \apj, 668, 853

\bibitem[{{Karim} {et~al.}(2011){Karim}, {Schinnerer},
  {Mart{\'{\i}}nez-Sansigre}, {Sargent}, {van der Wel}, {Rix}, {Ilbert},
  {Smol{\v c}i{\'c}}, {Carilli}, {Pannella}, {Koekemoer}, {Bell}, \&
  {Salvato}}]{karim11}
{Karim}, A., {et~al.} 2011, \apj, 730, 61

\bibitem[{{Kennicutt}(1998)}]{kennicutt98}
{Kennicutt}, Jr., R.~C. 1998, \araa, 36, 189

\bibitem[{{Kewley} {et~al.}(2001){Kewley}, {Dopita}, {Sutherland}, {Heisler},
  \& {Trevena}}]{kewley01}
{Kewley}, L.~J., {Dopita}, M.~A., {Sutherland}, R.~S., {Heisler}, C.~A., \&
  {Trevena}, J. 2001, \apj, 556, 121

\bibitem[{{Kewley} {et~al.}(2013){Kewley}, {Maier}, {Yabe}, {Ohta}, {Akiyama},
  {Dopita}, \& {Yuan}}]{kewley13}
{Kewley}, L.~J., {Maier}, C., {Yabe}, K., {Ohta}, K., {Akiyama}, M., {Dopita},
  M.~A., \& {Yuan}, T. 2013, \apjl, 774, L10

\bibitem[{{Koekemoer} {et~al.}(2011){Koekemoer}, {Faber}, {Ferguson}, {Grogin},
  {Kocevski}, {Koo}, {Lai}, {Lotz}, {Lucas}, {McGrath}, {Ogaz}, {Rajan},
  {Riess}, {Rodney}, {Strolger}, {Casertano}, {Castellano}, {Dahlen},
  {Dickinson}, {Dolch}, {Fontana}, {Giavalisco}, {Grazian}, {Guo}, {Hathi},
  {Huang}, {van der Wel}, {Yan}, {Acquaviva}, {Alexander}, {Almaini}, {Ashby},
  {Barden}, {Bell}, {Bournaud}, {Brown}, {Caputi}, {Cassata}, {Challis},
  {Chary}, {Cheung}, {Cirasuolo}, {Conselice}, {Roshan Cooray}, {Croton},
  {Daddi}, {Dav{\'e}}, {de Mello}, {de Ravel}, {Dekel}, {Donley}, {Dunlop},
  {Dutton}, {Elbaz}, {Fazio}, {Filippenko}, {Finkelstein}, {Frazer}, {Gardner},
  {Garnavich}, {Gawiser}, {Gruetzbauch}, {Hartley}, {H{\"a}ussler},
  {Herrington}, {Hopkins}, {Huang}, {Jha}, {Johnson}, {Kartaltepe},
  {Khostovan}, {Kirshner}, {Lani}, {Lee}, {Li}, {Madau}, {McCarthy},
  {McIntosh}, {McLure}, {McPartland}, {Mobasher}, {Moreira}, {Mortlock},
  {Moustakas}, {Mozena}, {Nandra}, {Newman}, {Nielsen}, {Niemi}, {Noeske},
  {Papovich}, {Pentericci}, {Pope}, {Primack}, {Ravindranath}, {Reddy},
  {Renzini}, {Rix}, {Robaina}, {Rosario}, {Rosati}, {Salimbeni}, {Scarlata},
  {Siana}, {Simard}, {Smidt}, {Snyder}, {Somerville}, {Spinrad}, {Straughn},
  {Telford}, {Teplitz}, {Trump}, {Vargas}, {Villforth}, {Wagner}, {Wandro},
  {Wechsler}, {Weiner}, {Wiklind}, {Wild}, {Wilson}, {Wuyts}, \&
  {Yun}}]{koekemoer11}
{Koekemoer}, A.~M., {et~al.} 2011, \apjs, 197, 36

\bibitem[{{Kriek} {et~al.}(2009){Kriek}, {van Dokkum}, {Labb{\'e}}, {Franx},
  {Illingworth}, {Marchesini}, \& {Quadri}}]{kriek09}
{Kriek}, M., {van Dokkum}, P.~G., {Labb{\'e}}, I., {Franx}, M., {Illingworth},
  G.~D., {Marchesini}, D., \& {Quadri}, R.~F. 2009, \apj, 700, 221

\bibitem[{{Kroupa}(2001)}]{kroupa01}
{Kroupa}, P. 2001, \mnras, 322, 231

\bibitem[{{Laird} {et~al.}(2009){Laird}, {Nandra}, {Georgakakis}, {Aird},
  {Barmby}, {Conselice}, {Coil}, {Davis}, {Faber}, {Fazio}, {Guhathakurta},
  {Koo}, {Sarajedini}, \& {Willmer}}]{laird09}
{Laird}, E.~S., {et~al.} 2009, \apjs, 180, 102

\bibitem[{{Lara-L{\'o}pez} {et~al.}(2013){Lara-L{\'o}pez}, {Hopkins},
  {L{\'o}pez-S{\'a}nchez}, {Brough}, {Colless}, {Bland-Hawthorn}, {Driver},
  {Foster}, {Liske}, {Loveday}, {Robotham}, {Sharp}, {Steele}, \&
  {Taylor}}]{lara-lopez13a}
{Lara-L{\'o}pez}, M.~A., {et~al.} 2013, \mnras, 433, L35

\bibitem[{{Lawrence} {et~al.}(2007){Lawrence}, {Warren}, {Almaini}, {Edge},
  {Hambly}, {Jameson}, {Lucas}, {Casali}, {Adamson}, {Dye}, {Emerson},
  {Foucaud}, {Hewett}, {Hirst}, {Hodgkin}, {Irwin}, {Lodieu}, {McMahon},
  {Simpson}, {Smail}, {Mortlock}, \& {Folger}}]{lawrence07}
{Lawrence}, A., {et~al.} 2007, \mnras, 379, 1599

\bibitem[{{Lee} {et~al.}(2007){Lee}, {Kennicutt}, {Funes}, {Sakai}, \&
  {Akiyama}}]{lee07}
{Lee}, J.~C., {Kennicutt}, R.~C., {Funes}, Jos{\'e}~G., S.~J., {Sakai}, S., \&
  {Akiyama}, S. 2007, \apjl, 671, L113

\bibitem[{{Lee} {et~al.}(2009{\natexlab{a}}){Lee}, {Kennicutt}, {Funes},
  {Sakai}, \& {Akiyama}}]{lee09}
{Lee}, J.~C., {Kennicutt}, Jr., R.~C., {Funes}, S.~J.~J.~G., {Sakai}, S., \&
  {Akiyama}, S. 2009{\natexlab{a}}, \apj, 692, 1305

\bibitem[{{Lee} {et~al.}(2009{\natexlab{b}}){Lee}, {Idzi}, {Ferguson},
  {Somerville}, {Wiklind}, \& {Giavalisco}}]{lees09}
{Lee}, S.-K., {Idzi}, R., {Ferguson}, H.~C., {Somerville}, R.~S., {Wiklind},
  T., \& {Giavalisco}, M. 2009{\natexlab{b}}, \apjs, 184, 100

\bibitem[{{Madau} {et~al.}(1996){Madau}, {Ferguson}, {Dickinson}, {Giavalisco},
  {Steidel}, \& {Fruchter}}]{madau96}
{Madau}, P., {Ferguson}, H.~C., {Dickinson}, M.~E., {Giavalisco}, M.,
  {Steidel}, C.~C., \& {Fruchter}, A. 1996, \mnras, 283, 1388

\bibitem[{{Masters} {et~al.}(2014){Masters}, {McCarthy}, {Siana}, {Malkan},
  {Mobasher}, {Atek}, {Henry}, {Martin}, {Rafelski}, {Hathi}, {Scarlata},
  {Ross}, {Bunker}, {Blanc}, {Bedregal}, {Dominguez}, {Colbert}, {Teplitz}, \&
  {Dressler}}]{masters14}
{Masters}, D., {et~al.} 2014, ArXiv e-prints

\bibitem[{{McLure} {et~al.}(2011){McLure}, {Dunlop}, {de Ravel}, {Cirasuolo},
  {Ellis}, {Schenker}, {Robertson}, {Koekemoer}, {Stark}, \&
  {Bowler}}]{mclure11}
{McLure}, R.~J., {et~al.} 2011, ArXiv e-prints

\bibitem[{{Momcheva} {et~al.}(2013){Momcheva}, {Lee}, {Ly}, {Salim}, {Dale},
  {Ouchi}, {Finn}, \& {Ono}}]{momcheva13}
{Momcheva}, I.~G., {Lee}, J.~C., {Ly}, C., {Salim}, S., {Dale}, D.~A., {Ouchi},
  M., {Finn}, R., \& {Ono}, Y. 2013, \aj, 145, 47

\bibitem[{{Newman} {et~al.}(2012){Newman}, {Cooper}, {Davis}, {Faber}, {Coil},
  {Guhathakurta}, {Koo}, {Phillips}, {Conroy}, {Dutton}, {Finkbeiner}, {Gerke},
  {Rosario}, {Weiner}, {Willmer}, {Yan}, {Harker}, {Kassin}, {Konidaris},
  {Lai}, {Madgwick}, {Noeske}, {Wirth}, {Connolly}, {Kaiser}, {Kirby},
  {Lemaux}, {Lin}, {Lotz}, {Luppino}, {Marinoni}, {Matthews}, {Metevier}, \&
  {Schiavon}}]{newman12}
{Newman}, J.~A., {et~al.} 2012, ArXiv e-prints

\bibitem[{{Noeske} {et~al.}(2007){Noeske}, {Weiner}, {Faber}, {Papovich},
  {Koo}, {Somerville}, {Bundy}, {Conselice}, {Newman}, {Schiminovich}, {Le
  Floc'h}, {Coil}, {Rieke}, {Lotz}, {Primack}, {Barmby}, {Cooper}, {Davis},
  {Ellis}, {Fazio}, {Guhathakurta}, {Huang}, {Kassin}, {Martin}, {Phillips},
  {Rich}, {Small}, {Willmer}, \& {Wilson}}]{noeske07}
{Noeske}, K.~G., {et~al.} 2007, \apjl, 660, L43

\bibitem[{{Ono} {et~al.}(2010){Ono}, {Ouchi}, {Shimasaku}, {Dunlop}, {Farrah},
  {McLure}, \& {Okamura}}]{ono10}
{Ono}, Y., {Ouchi}, M., {Shimasaku}, K., {Dunlop}, J., {Farrah}, D., {McLure},
  R., \& {Okamura}, S. 2010, \apj, 724, 1524

\bibitem[{{Osterbrock}(1989)}]{osterbrock89}
{Osterbrock}, D.~E. 1989, {Astrophysics of gaseous nebulae and active galactic
  nuclei} (Research supported by the University of California, John Simon
  Guggenheim Memorial Foundation, University of Minnesota, et al.~Mill Valley,
  CA, University Science Books, 1989, 422 p.)

\bibitem[{{Pacifici} {et~al.}(2012){Pacifici}, {Charlot}, {Blaizot}, \&
  {Brinchmann}}]{pacifici12}
{Pacifici}, C., {Charlot}, S., {Blaizot}, J., \& {Brinchmann}, J. 2012, \mnras,
  421, 2002

\bibitem[{{Pacifici} {et~al.}(2013){Pacifici}, {Kassin}, {Weiner}, {Charlot},
  \& {Gardner}}]{pacifici13}
{Pacifici}, C., {Kassin}, S.~A., {Weiner}, B., {Charlot}, S., \& {Gardner},
  J.~P. 2013, \apjl, 762, L15

\bibitem[{{Pannella} {et~al.}(2009){Pannella}, {Carilli}, {Daddi}, {McCracken},
  {Owen}, {Renzini}, {Strazzullo}, {Civano}, {Koekemoer}, {Schinnerer},
  {Scoville}, {Smol{\v c}i{\'c}}, {Taniguchi}, {Aussel}, {Kneib}, {Ilbert},
  {Mellier}, {Salvato}, {Thompson}, \& {Willott}}]{pannella09}
{Pannella}, M., {et~al.} 2009, \apjl, 698, L116

\bibitem[{{Peng} {et~al.}(2010){Peng}, {Lilly}, {Kova{\v c}}, {Bolzonella},
  {Pozzetti}, {Renzini}, {Zamorani}, {Ilbert}, {Knobel}, {Iovino}, {Maier},
  {Cucciati}, {Tasca}, {Carollo}, {Silverman}, {Kampczyk}, {de Ravel},
  {Sanders}, {Scoville}, {Contini}, {Mainieri}, {Scodeggio}, {Kneib}, {Le
  F{\`e}vre}, {Bardelli}, {Bongiorno}, {Caputi}, {Coppa}, {de la Torre},
  {Franzetti}, {Garilli}, {Lamareille}, {Le Borgne}, {Le Brun}, {Mignoli},
  {Perez Montero}, {Pello}, {Ricciardelli}, {Tanaka}, {Tresse}, {Vergani},
  {Welikala}, {Zucca}, {Oesch}, {Abbas}, {Barnes}, {Bordoloi}, {Bottini},
  {Cappi}, {Cassata}, {Cimatti}, {Fumana}, {Hasinger}, {Koekemoer},
  {Leauthaud}, {Maccagni}, {Marinoni}, {McCracken}, {Memeo}, {Meneux}, {Nair},
  {Porciani}, {Presotto}, \& {Scaramella}}]{peng10}
{Peng}, Y.-j., {et~al.} 2010, \apj, 721, 193

\bibitem[{{Reddy} {et~al.}(2010){Reddy}, {Erb}, {Pettini}, {Steidel}, \&
  {Shapley}}]{reddy10}
{Reddy}, N.~A., {Erb}, D.~K., {Pettini}, M., {Steidel}, C.~C., \& {Shapley},
  A.~E. 2010, \apj, 712, 1070

\bibitem[{{Rodighiero} {et~al.}(2010){Rodighiero}, {Cimatti}, {Gruppioni},
  {Popesso}, {Andreani}, {Altieri}, {Aussel}, {Berta}, {Bongiovanni},
  {Brisbin}, {Cava}, {Cepa}, {Daddi}, {Dominguez-Sanchez}, {Elbaz}, {Fontana},
  {F{\"o}rster Schreiber}, {Franceschini}, {Genzel}, {Grazian}, {Lutz},
  {Magdis}, {Magliocchetti}, {Magnelli}, {Maiolino}, {Mancini}, {Nordon},
  {Perez Garcia}, {Poglitsch}, {Santini}, {Sanchez-Portal}, {Pozzi},
  {Riguccini}, {Saintonge}, {Shao}, {Sturm}, {Tacconi}, {Valtchanov},
  {Wetzstein}, \& {Wieprecht}}]{rodighiero10}
{Rodighiero}, G., {et~al.} 2010, \aap, 518, L25

\bibitem[{{Rodighiero} {et~al.}(2011){Rodighiero}, {Daddi}, {Baronchelli},
  {Cimatti}, {Renzini}, {Aussel}, {Popesso}, {Lutz}, {Andreani}, {Berta},
  {Cava}, {Elbaz}, {Feltre}, {Fontana}, {F{\"o}rster Schreiber},
  {Franceschini}, {Genzel}, {Grazian}, {Gruppioni}, {Ilbert}, {Le Floch},
  {Magdis}, {Magliocchetti}, {Magnelli}, {Maiolino}, {McCracken}, {Nordon},
  {Poglitsch}, {Santini}, {Pozzi}, {Riguccini}, {Tacconi}, {Wuyts}, \&
  {Zamorani}}]{rodighiero11}
---. 2011, \apjl, 739, L40

\bibitem[{{Salmi} {et~al.}(2012){Salmi}, {Daddi}, {Elbaz}, {Sargent},
  {Dickinson}, {Renzini}, {Bethermin}, \& {Le Borgne}}]{salmi12}
{Salmi}, F., {Daddi}, E., {Elbaz}, D., {Sargent}, M.~T., {Dickinson}, M.,
  {Renzini}, A., {Bethermin}, M., \& {Le Borgne}, D. 2012, \apjl, 754, L14

\bibitem[{{Salpeter}(1955)}]{salpeter55}
{Salpeter}, E.~E. 1955, \apj, 121, 161

\bibitem[{{Savaglio} {et~al.}(2005){Savaglio}, {Glazebrook}, {Le Borgne},
  {Juneau}, {Abraham}, {Chen}, {Crampton}, {McCarthy}, {Carlberg}, {Marzke},
  {Roth}, {J{\o}rgensen}, \& {Murowinski}}]{savaglio05}
{Savaglio}, S., {et~al.} 2005, \apj, 635, 260

\bibitem[{{Scalo}(1986)}]{scalo86}
{Scalo}, J.~M. 1986, \fcp, 11, 1

\bibitem[{{Schaerer} \& {de Barros}(2009)}]{schaerer09}
{Schaerer}, D., \& {de Barros}, S. 2009, \aap, 502, 423

\bibitem[{{Schaerer} {et~al.}(2011){Schaerer}, {de Barros}, \&
  {Stark}}]{schaerer11}
{Schaerer}, D., {de Barros}, S., \& {Stark}, D.~P. 2011, \aap, 536, A72

\bibitem[{{Schenker} {et~al.}(2013){Schenker}, {Ellis}, {Konidaris}, \&
  {Stark}}]{schenker13}
{Schenker}, M.~A., {Ellis}, R.~S., {Konidaris}, N.~P., \& {Stark}, D.~P. 2013,
  ArXiv e-prints

\bibitem[{{Shen} {et~al.}(2013){Shen}, {Madau}, {Conroy}, {Governato}, \&
  {Mayer}}]{shen13}
{Shen}, S., {Madau}, P., {Conroy}, C., {Governato}, F., \& {Mayer}, L. 2013,
  ArXiv e-prints

\bibitem[{{Shim} {et~al.}(2011){Shim}, {Chary}, {Dickinson}, {Lin}, {Spinrad},
  {Stern}, \& {Yan}}]{shim11}
{Shim}, H., {Chary}, R.-R., {Dickinson}, M., {Lin}, L., {Spinrad}, H., {Stern},
  D., \& {Yan}, C.-H. 2011, ArXiv e-prints

\bibitem[{{Silk}(2013)}]{silk13}
{Silk}, J. 2013, \apj, 772, 112

\bibitem[{{Sobral} {et~al.}(2012){Sobral}, {Best}, {Matsuda}, {Smail}, {Geach},
  \& {Cirasuolo}}]{sobral12}
{Sobral}, D., {Best}, P.~N., {Matsuda}, Y., {Smail}, I., {Geach}, J.~E., \&
  {Cirasuolo}, M. 2012, \mnras, 420, 1926

\bibitem[{{Springel} {et~al.}(2005){Springel}, {White}, {Jenkins}, {Frenk},
  {Yoshida}, {Gao}, {Navarro}, {Thacker}, {Croton}, {Helly}, {Peacock}, {Cole},
  {Thomas}, {Couchman}, {Evrard}, {Colberg}, \& {Pearce}}]{springel05}
{Springel}, V., {et~al.} 2005, \nat, 435, 629

\bibitem[{{Taniguchi} {et~al.}(2010){Taniguchi}, {Shioya}, \&
  {Trump}}]{taniguchi10}
{Taniguchi}, Y., {Shioya}, Y., \& {Trump}, J.~R. 2010, \apj, 724, 1480

\bibitem[{{Tremonti} {et~al.}(2004){Tremonti}, {Heckman}, {Kauffmann},
  {Brinchmann}, {Charlot}, {White}, {Seibert}, {Peng}, {Schlegel}, {Uomoto},
  {Fukugita}, \& {Brinkmann}}]{tremonti04}
{Tremonti}, C.~A., {et~al.} 2004, \apj, 613, 898

\bibitem[{{Ueda} {et~al.}(2008){Ueda}, {Watson}, {Stewart}, {Akiyama},
  {Schwope}, {Lamer}, {Ebrero}, {Carrera}, {Sekiguchi}, {Yamada}, {Simpson},
  {Hasinger}, \& {Mateos}}]{ueda08}
{Ueda}, Y., {et~al.} 2008, \apjs, 179, 124

\bibitem[{{Watson} {et~al.}(2010){Watson}, {French}, {Christensen},
  {O'Halloran}, {Micha{\l}owski}, {Hjorth}, {Malesani}, {Fynbo}, {Gordon}, \&
  {Castro Cer{\'o}n}}]{watson10}
{Watson}, D., {et~al.} 2010, ArXiv e-prints

\bibitem[{{Werner} {et~al.}(2004){Werner}, {Roellig}, {Low}, {Rieke}, {Rieke},
  {Hoffmann}, {Young}, {Houck}, {Brandl}, {Fazio}, {Hora}, {Gehrz}, {Helou},
  {Soifer}, {Stauffer}, {Keene}, {Eisenhardt}, {Gallagher}, {Gautier}, {Irace},
  {Lawrence}, {Simmons}, {Van Cleve}, {Jura}, {Wright}, \&
  {Cruikshank}}]{werner04}
{Werner}, M.~W., {et~al.} 2004, \apjs, 154, 1

\bibitem[{{Whitaker} {et~al.}(2012){Whitaker}, {van Dokkum}, {Brammer}, \&
  {Franx}}]{whitaker12}
{Whitaker}, K.~E., {van Dokkum}, P.~G., {Brammer}, G., \& {Franx}, M. 2012,
  \apjl, 754, L29

\bibitem[{{Whitaker} {et~al.}(2011){Whitaker}, {Labb{\'e}}, {van Dokkum},
  {Brammer}, {Kriek}, {Marchesini}, {Quadri}, {Franx}, {Muzzin}, {Williams},
  {Bezanson}, {Illingworth}, {Lee}, {Lundgren}, {Nelson}, {Rudnick}, {Tal}, \&
  {Wake}}]{whitaker11}
{Whitaker}, K.~E., {et~al.} 2011, \apj, 735, 86

\bibitem[{{Williams} {et~al.}(2011){Williams}, {Quadri}, \&
  {Franx}}]{williams11}
{Williams}, R.~J., {Quadri}, R.~F., \& {Franx}, M. 2011, \apjl, 738, L25

\bibitem[{{Wuyts} {et~al.}(2008){Wuyts}, {Labb{\'e}}, {Schreiber}, {Franx},
  {Rudnick}, {Brammer}, \& {van Dokkum}}]{wuyts08}
{Wuyts}, S., {Labb{\'e}}, I., {Schreiber}, N.~M.~F., {Franx}, M., {Rudnick},
  G., {Brammer}, G.~B., \& {van Dokkum}, P.~G. 2008, \apj, 682, 985

\bibitem[{{Wuyts} {et~al.}(2011){Wuyts}, {F{\"o}rster Schreiber}, {van der
  Wel}, {Magnelli}, {Guo}, {Genzel}, {Lutz}, {Aussel}, {Barro}, {Berta},
  {Cava}, {Graci{\'a}-Carpio}, {Hathi}, {Huang}, {Kocevski}, {Koekemoer},
  {Lee}, {Le Floc'h}, {McGrath}, {Nordon}, {Popesso}, {Pozzi}, {Riguccini},
  {Rodighiero}, {Saintonge}, \& {Tacconi}}]{wuyts11}
{Wuyts}, S., {et~al.} 2011, \apj, 742, 96

\bibitem[{{Xue} {et~al.}(2011){Xue}, {Luo}, {Brandt}, {Bauer}, {Lehmer},
  {Broos}, {Schneider}, {Alexander}, {Brusa}, {Comastri}, {Fabian}, {Gilli},
  {Hasinger}, {Hornschemeier}, {Koekemoer}, {Liu}, {Mainieri}, {Paolillo},
  {Rafferty}, {Rosati}, {Shemmer}, {Silverman}, {Smail}, {Tozzi}, \&
  {Vignali}}]{xue11}
{Xue}, Y.~Q., {et~al.} 2011, \apjs, 195, 10

\end{thebibliography}

\end{document}